\documentclass{aa}
\usepackage[bookmarks,bookmarksopen,colorlinks,linkcolor={blue},citecolor={blue},
  urlcolor={blue}]{hyperref}
\usepackage{amsmath}
\usepackage{natbib}
\usepackage{color}
\usepackage[x11names]{xcolor}
\usepackage{graphics}
\usepackage{geometry}
\usepackage{enumerate}
\usepackage{setspace}
\usepackage{siunitx}
\usepackage{pdflscape}

\newcommand{\mps}[1]{m s$^{-1}$#1}

\newcommand{\prot}[1]{$P_{\text{rot}}$#1}
\defcitealias{cloutier17b}{C17a}
\defcitealias{sarkis18}{S18}

\newcommand{\cdbox}[1]{%
  \colorlet{currentcolor}{.}%
  {\color{Blue1}%
    \dbox{\color{currentcolor}#1}}%
}
\usepackage{dashbox,framed,color,ocg-p}
\newcommand{\ToggleLayer}[2]{%
  \leavevmode
  \pdfstartlink user {
    /Subtype /Link
    /Border [0 0 0]%
    /A <<
      /S/JavaScript
      /JS (
         var aOCGs = this.getOCGs(), Layer;
         var Layers = "#1".split(","), Active = -1, i, l;
         for (l=0; l<Layers.length; l++) {
           Layer = Layers[l];
           for (i=0; aOCGs && i<aOCGs.length; i++) {
             if (aOCGs[i].state && aOCGs[i].name == Layer) {
               Active = l;
               aOCGs[i].state = false;
             }
           }
           if (Active >= 0) break;
         }
         if (Active == -1) {
           for (l=0; l<Layers.length; l++) {
             if (Layers[l] == "") Active = l;
           }
         }
         Active = Active + 1;
         if (Active == Layers.length) Active = 0;
         Layer = Layers[Active];
         for (i=0; aOCGs && i<aOCGs.length; i++) {
           if (aOCGs[i].name == Layer) aOCGs[i].state = true;
         }
      )
    >>
  }#2%
  \pdfendlink
}

\begin{document}

\title{Confirmation of the radial velocity super-Earth K2-18c with HARPS\thanks{Based on observations made with the HARPS instrument on the ESO 3.6 m telescope under the program IDs 191.C-0873(A), and 198.C-0838(A) at Cerro La Silla (Chile). Radial velocity data will be available in electronic form at the CDS via anonymous ftp to cdsarc.u-strasbg.fr (130.79.128.5) or via http://cdsweb.u-strasbg.fr/cgi-bin/qcat?J/A+A/} and CARMENES}
\titlerunning{Confirmation of the RV planet K2-18c}
\authorrunning{Cloutier et al.:}

\author{R.~Cloutier \inst{1,2,3}
  \and N.~Astudillo-Defru \inst{4}
  \and R.~Doyon \inst{3}
  \and X.~Bonfils \inst{5}
  \and J.-M.~Almenara \inst{6}
  \and F.~Bouchy \inst{6}
  \and X.~Delfosse \inst{5}
  \and T.~Forveille \inst{5}
  \and C.~Lovis \inst{6}
  \and M.~Mayor \inst{6}
  \and K.~Menou \inst{1,2}
  \and F.~Murgas \inst{5}
  \and F.~Pepe \inst{6}
  \and N.~C.~Santos \inst{7,8}
  \and S.~Udry \inst{6}
  \and A.~W\"unsche \inst{5}
}

\institute{Dept. of Astronomy \& Astrophysics, University of Toronto, 50 St. George Street, M5S 3H4, Toronto, ON, Canada
  \email{cloutier@astro.utoronto.ca}
  \and Centre for Planetary Sciences, Dept. of Physical \& Environmental Sciences, University of Toronto Scarborough, 1265 Military Trail, M1C 1A4, Toronto, ON, Canada
  \and Institut de Recherche sur les Exoplan\`etes, d\'epartement de physique, Universit\'e de Montr\'eal, C.P. 6128 Succ. Centre-ville, H3C 3J7, Montr\'eal, QC, Canada
  \and Universidad de Concepci\'on, Departamento de Astronom\'ia, Casilla 160-C, Concepci\'on, Chile
  \and Universit\'e Grenoble Alpes, CNRS, IPAG, F-38000 Grenoble, France
  \and Observatoire Astronomique de l’Universit\'e de Gen\`eve, 51 chemin des Maillettes, 1290 Versoix, Switzerland
  \and Instituto de Astrof\'isica e Ci\^encias do Espa\c{c}o, Universidade do Porto, CAUP, Rua das Estrelas, 4150-762 Porto, Portugal
  \and Departamento de F\'isica e Astronomia, Faculdade de Ci\^encias, Universidade do Porto, Rua do Campo Alegre, 4169-007 Porto, Portugal
}

\abstract{In an earlier campaign to characterize the mass of the transiting temperate super-Earth
  K2-18b with HARPS, a second, non-transiting planet was
  posited to exist in the system at $\sim 9$ days. Further radial velocity follow-up with the
  CARMENES spectrograph visible channel revealed
  a much weaker signal at 9 days which also appeared to vary chromatically and temporally
  leading to the conclusion that the origin of the 9 day signal was more likely to be related to
  stellar activity than to being planetary. Here we conduct a detailed re-analysis of all available
  RV time-series---including a set of 31 previously unpublished HARPS measurements---to investigate
  the effects of time-sampling and of simultaneous modelling of
  planetary + activity signals on the existence and origin of the curious 9 day signal. We conclude
  that the 9 day signal is real and was initially seen to be suppressed in the CARMENES data
  due to a small number of anomalous measurements although the exact cause of these anomalies remains unknown.
  Investigation of the signal's evolution in time, with wavelength, and detailed model comparison reveals
  that the 9 day signal is most likely planetary in nature. By this analysis we reconcile the conflicting HARPS
  and CARMENES results and measure precise and self-consistent planet masses of
  $m_{p,b} = 8.63 \pm 1.35$ and $m_{p,c}\sin{i_c}=5.62 \pm 0.84$ Earth masses. This work---along with the
  previously published RV papers on the K2-18 planetary system---highlight the importance of understanding
  one's time-sampling and of simultaneous planet + stochastic activity modelling, particularly when searching
  for sub-Neptune-sized planets with radial velocities.}

\maketitle

\section{Introduction}
The nearby M2.5 dwarf K2-18 (EPIC 201912552, $d\sim 38$ pc, $J=9.8$)
is known to host a transiting sub-Neptune-sized planet 
at $\sim 33$ days; K2-18b \citep{foremanmackey15b,montet15,benneke17}. Given the planet's orbital separation
and corresponding equilibrium temperature, K2-18b is a temperate planet and represents one of the
most attractive targets for the atmospheric characterization of a habitable zone exoplanet 
that was discovered in the pre-TESS era. Indeed K2-18b is already slated for transmission
spectroscopy observations as part of the NIRISS GTO program
1201\footnote{\url{http://www.stsci.edu/cgi-bin/get-proposal-info?id=1201&observatory=JWST}}.

Given the requirement for a-priori knowledge of a planet's bulk density in order to interpret
observations of its atmosphere, multiple groups have endeavored to
measure the mass of K2-18b via ground-based radial velocity (RV) measurements in the visible
wavelength domain. Specifically, \cite{cloutier17b} (hereafter \citetalias{cloutier17b})
first reported the mass of K2-18b to be
$8.0\pm 1.9$ M$_{\oplus}$ based on 75 measurements taken with the HARPS spectrograph on the ESO 3.6m
telescope at La Silla  \citep{mayor03}. 
Their RV time-series also exhibited a strong additional signal
at $\sim 9$ days which was not seen in any other contemporaneous activity indicator\footnote{e.g. the
  S-index, H$\alpha$ index, full width at half maximum, and the bi-sector inverse slope of the spectral
  cross-correlation function} nor in the window function.
\citetalias{cloutier17b} presented evidence for the
planetary nature of the 9 day signal by simultaneously modelling both planetary signals with keplerians
and the correlated RV residuals using a trained quasi-periodic Gaussian process.
Correlated RV residuals--after the removal of planetary signals---are expected to arise from
stellar activity whose components can be seen in various activity indicators such as
photometry and the aforementioned spectroscopic indicators. All of these ancillary time-series were used
for training in the multiple analyzes presented in \citetalias{cloutier17b}.
Stellar activity on M dwarfs is largely modulated by stellar rotation \citep{boisse11}
and thus produces a \emph{quasi}-periodic structure in the RVs that is physically motivated.
Such correlated structure is often not strictly sinusoidal as the
active regions that give rise to the observed stellar activity have finite lifetimes, spatial
distributions, and temperature contrasts that evolve temporally over a
few rotation cycles and thus lead to non-sinusoidal structure over the observational baseline.

Recently, \cite{sarkis18} (hereafter \citetalias{sarkis18})
presented an independent set of 58 RV measurements of K2-18 taken with the visible
channel on CARMENES \citep[561-905 nm;][]{quirrenbach14}.
With these data \citetalias{sarkis18} independently measured the
mass of K2-18b to be $8.9^{+1.7}_{-1.6}$ M$_{\oplus}$, a result that is consistent with the measured value from
\citetalias{cloutier17b}. However in their data---with comparable RV precision---the 9 day signal with its
proposed planetary origin from \citetalias{cloutier17b} was only marginally detected. Furthermore,
\citetalias{sarkis18} claimed that the signal was seen to vary in time and whose strength (as measured by
the false alarm probability in the generalized Lomb-Scargle periodogram) appeared to vary with
wavelength. Given the proximity of the 9 day signal to the fourth harmonic of the photometric stellar
rotation period\footnote{Although periodicities at the second and third harmonics are not seen in the CARMENES
  RVs with comparable significance to that of the 9 day signal.} \citepalias[\prot{} $=38.6$ days;][]{cloutier17b}, 
\citetalias{sarkis18} interpreted the weak 9 day signal as one whose origin is more
likely due to stellar activity than to a second, non-transiting planet in the system. 

Based on the strong evidence for the detection of K2-18c with HARPS\footnote{i.e. a strong
  periodic signal in the periodogram of the HARPS RVs at $\sim 9$ days, a $6.3\sigma$ semi-amplitude measurement,
  the favourability of a 2-planet model by cross-validation model comparison \citepalias{cloutier17b}.} and the
low significance of its periodic signal being seen with CARMENES, here we conduct a systematic re-analysis of all 
available RV data to confirm or disprove the existence of a stable periodic signal at $\sim 9$ days in the K2-18
system and ultimately to determine the nature of that signal as planetary or otherwise. In this study we
independently analyze the aforementioned HARPS and CARMENES RV time-series and their joint time-series.
We include 31 previously unpublished HARPS RVs that aid in the interpretation
of the 9 day signal and improve the measurement precision of the planetary parameters.
In Sect.~\ref{sect:sampling} we present a detailed analysis investigating the effects of
time-sampling on the probability of the 9 day signal.
In Sects.~\ref{sect:act} and~\ref{sect:temporal} we investigate the proposed chromatic and temporal dependencies
of the 9 day signal with HARPS. In Sect.~\ref{sect:correlated} we self-consistently analyze all RVs in
the presence of a probabilistic correlated noise (i.e. activity) model.
Overall we find evidence for the planetary nature of the 9 day signal and conclude with a discussion in
Sect.~\ref{sect:dis}.

\section{The issue of sub-optimal window functions} \label{sect:sampling}
One potential reason for the strong 9 day signal to be seen in the published HARPS RVs and not with CARMENES may
be due to sub-optimal time-sampling  (i.e. the window function; WF). For example, the 9 day signal seen
with HARPS may arise from a sub-optimal WF and is therefore not associated with an astrophysical source such as a
planet or stellar activity. Similarly, if the 9 day signal exists and whose origin is physical then it is possible
that the CARMENES WF may suppress its signal in a Lomb-Scargle periodogram.
Indeed sub-optimal WFs have been shown to lead to inaccurate RV planet
masses and false planet detections (e.g. GL 581d; \citealt{hatzes16}, $\alpha$ Cen Bb; \citealt{rajpaul16},
Kepler-10c; \citealt{rajpaul17}).
Before proceeding we note that neither of the aforementioned scenarios are
expected to \emph{significantly} enhance or suppress the 9 day signal as investigated by preliminary analyses in
\citetalias{cloutier17b} and \citetalias{sarkis18}. However, a more subtle effect may be at play here. 
Specifically, the periodogram of the HARPS WF showed no excess
power at 9 days \citepalias[c.f. Fig. 2][]{cloutier17b} such that that signal is unlikely to originate from
sub-optimal HARPS sampling. Similarly,
\citetalias{sarkis18} created a synthetic RV time-series with the maximum a-posteriori (MAP) solution for
K2-18c from \citetalias{cloutier17b}---plus white noise---and sampled the keplerian curve with synthetic RVs using
the CARMENES WF. They reported that the $\sim 9$ day signal was seen in the periodogram
and thus was not suppressed by the CARMENES WF. Here we extend these analyses to establish
definitively whether or not either published WF is responsible for the ambiguity of the $\sim 9$ day signal.

Here we aim to establish the ease with which the
K2-18c signal at $\sim 9$ days can be detected in any of the published HARPS,
CARMENES, or joint WFs. Firstly, for each of the three WFs we construct a set of synthetic RV
time-series containing a variety of injected physical signals,
plus a white noise term with standard deviation equal to the mean
RV measurement precision of that time-series.\footnote{i.e. 3.60, 3.08, and 3.37 \mps{} for HARPS, CARMENES, and
  their joint time-series respectively.} We consider four flavors of injected physical signals of increasing
complexity: i) K2-18c only ii) K2-18b and c, iii) both planets plus correlated noise
due to stellar activity, and iv) K2-18b and stellar activity. The last time-series---which does not contain an injected
K2-18c signal---is included to test the hypothesis that the $P_c$ signal could arise without K2-18c existing at $P_c$ due
to sampling or stellar activity as posited by \citetalias{sarkis18}.
The test with K2-18b and c only using the CARMENES WF corresponds
to the test performed by \citetalias{sarkis18} which showed that $P_c$ is detected 
when the MAP value of the K2-18c semi-amplitude $K_c=4.63$ \mps{} from \citetalias{cloutier17b} was injected.
In our analysis, the keplerian model parameters
for each planet are fixed to their average value between the \citetalias{cloutier17b} and \citetalias{sarkis18}
results---where applicable---with the exception of $K_c$ which is sampled on a
logarithmically equidistant grid from 1-10 \mps{.} When including correlated noise models,
those models are sampled from a quasi-periodic Gaussian process prior distribution which has been shown to be
an effective means of describing quasi-periodic stellar activity signals in both Sun-like and M dwarf stars
\citep[e.g.][]{haywood14,cloutier17a}. The adopted hyperparameters are given by those measured in \emph{Model 1} from
\citetalias{cloutier17b} and includes a covariance amplitude of 2.8 \mps{.} These
hyperparameters describe the covariance structure of the stellar activity signal as seen in the star's K2 photometry
and the HARPS RVs.

\begin{figure*}
  \centering
  \includegraphics[width=0.65\hsize]{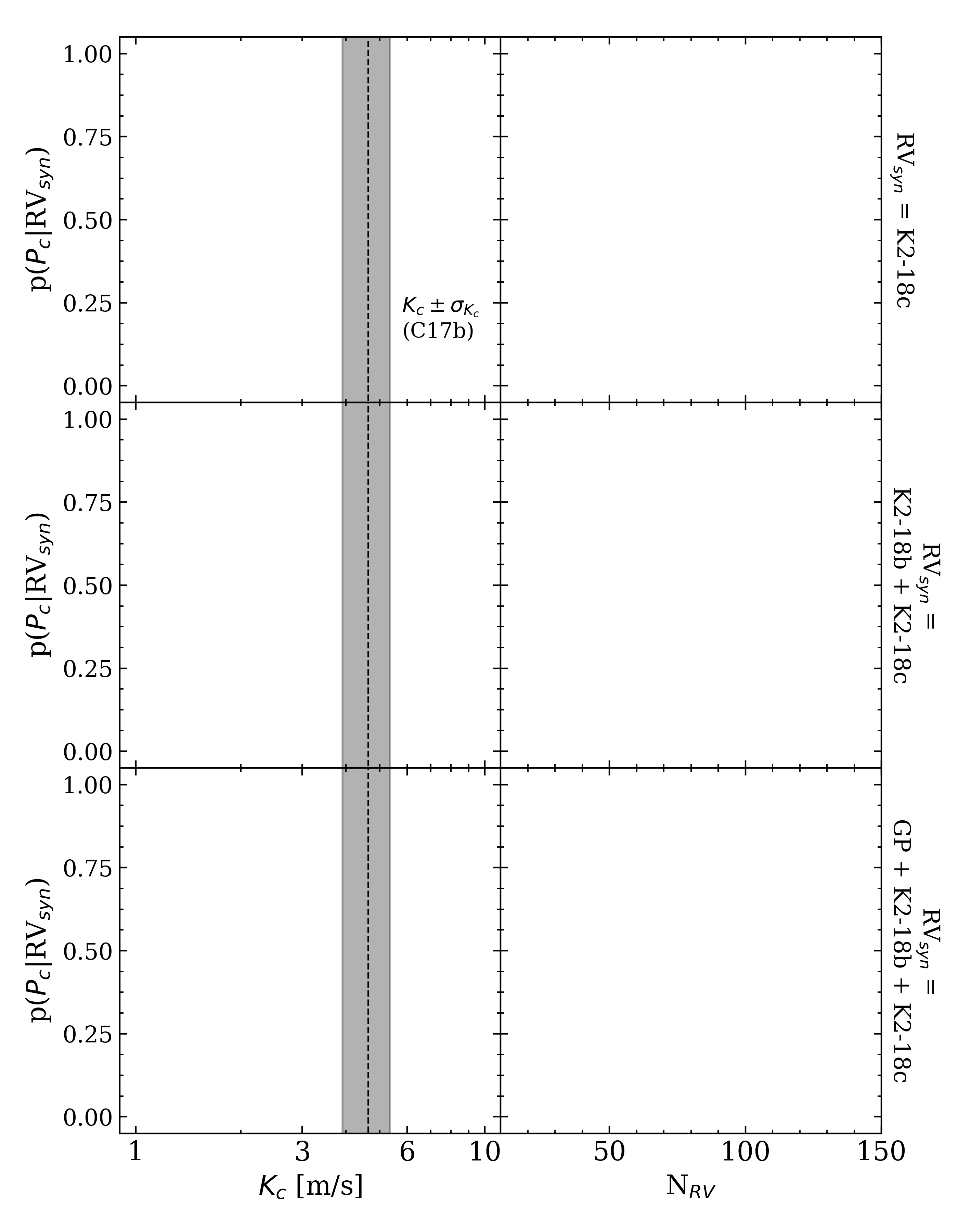}%
  \hspace{-0.65\hsize}%
  \begin{ocg}{fig:Hoff}{fig:Hoff}{0}%
  \end{ocg}%
  \begin{ocg}{fig:Hon}{fig:Hon}{1}%
  \includegraphics[width=0.65\hsize]{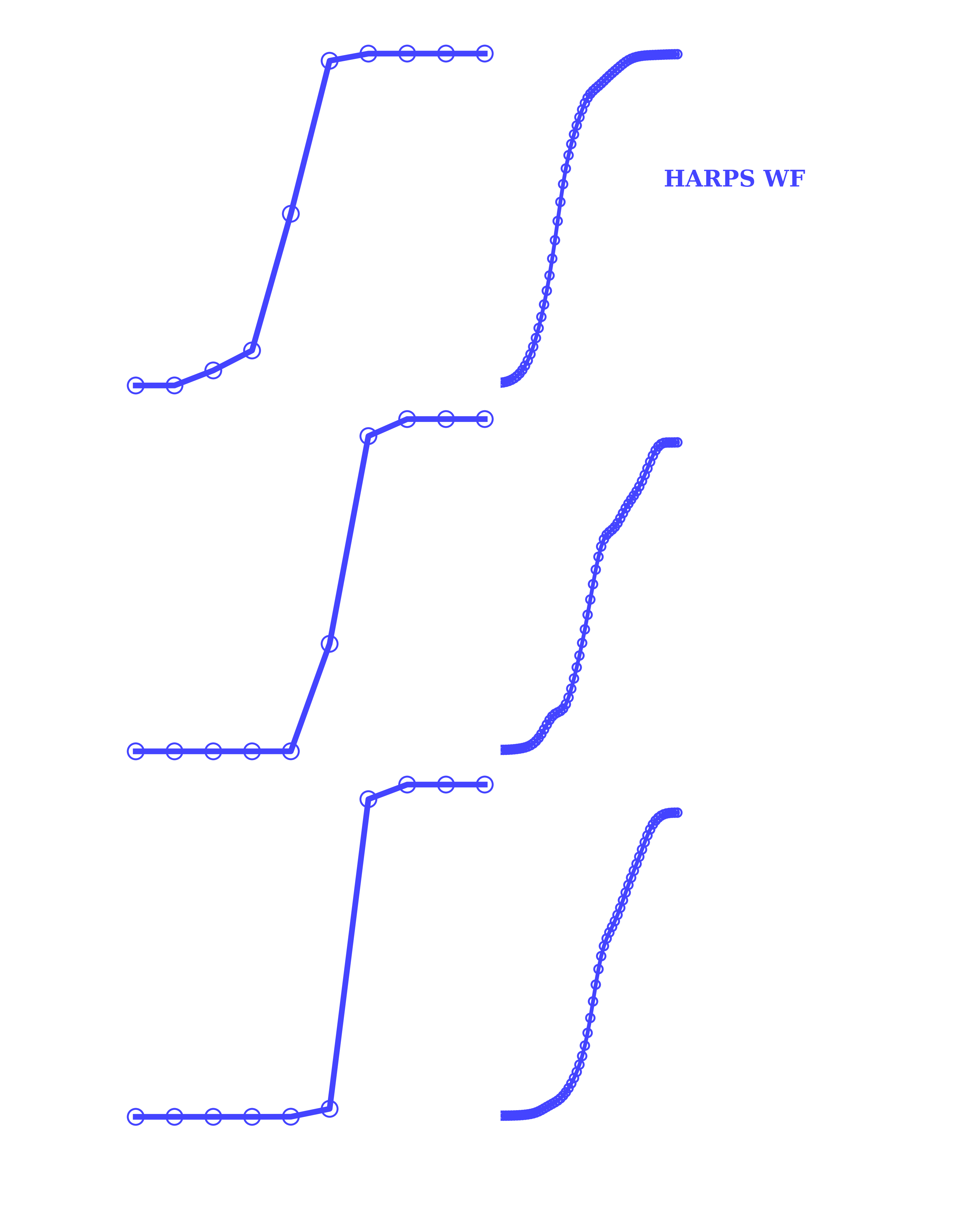}%
  \end{ocg}
  \hspace{-0.65\hsize}%
  \begin{ocg}{fig:Coff}{fig:Coff}{0}%
  \end{ocg}%
  \begin{ocg}{fig:Con}{fig:Con}{1}%
  \includegraphics[width=0.65\hsize]{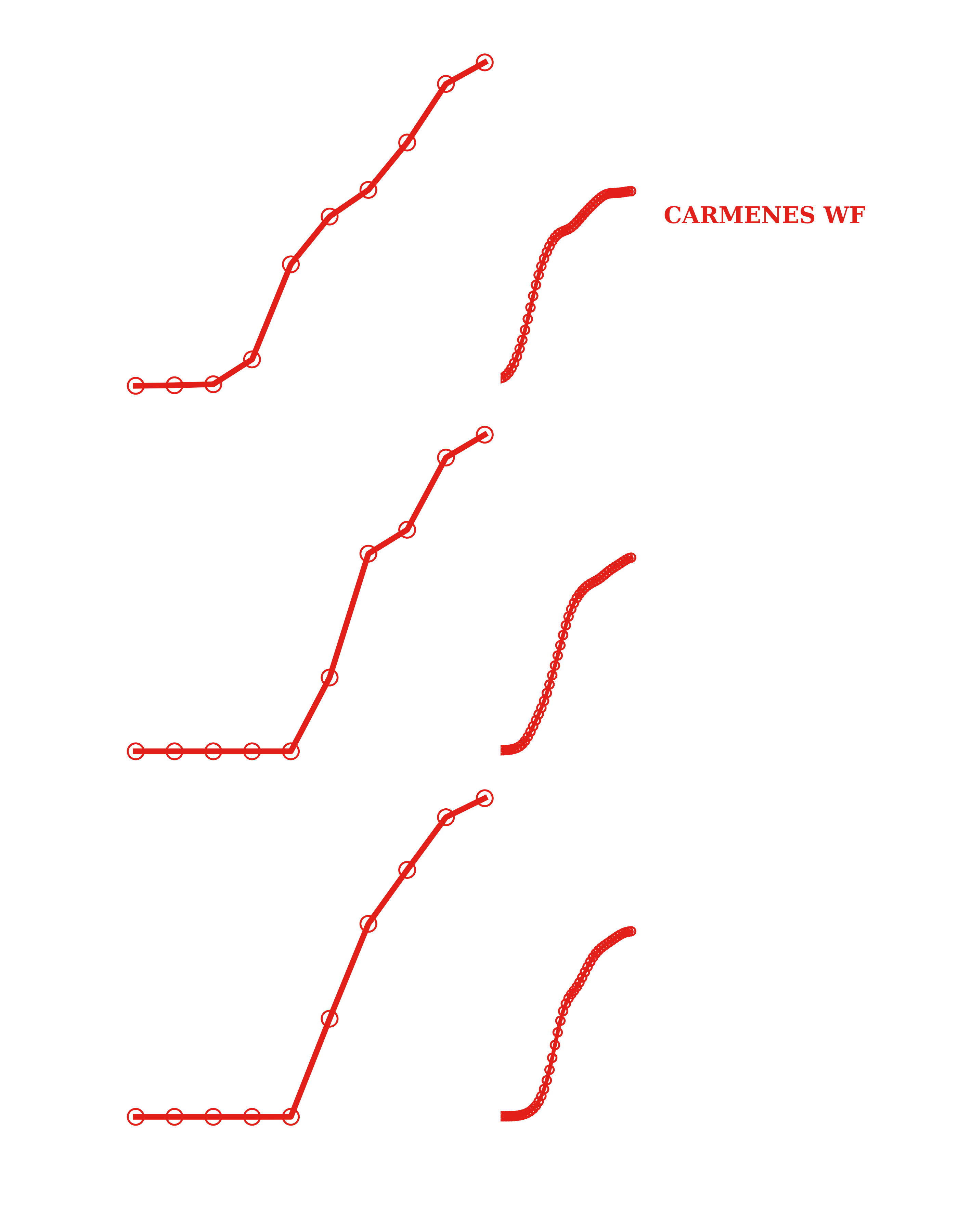}%
  \end{ocg}
  \hspace{-0.65\hsize}%
  \begin{ocg}{fig:joff}{fig:joff}{0}%
  \end{ocg}%
  \begin{ocg}{fig:jon}{fig:jon}{1}%
  \includegraphics[width=0.65\hsize]{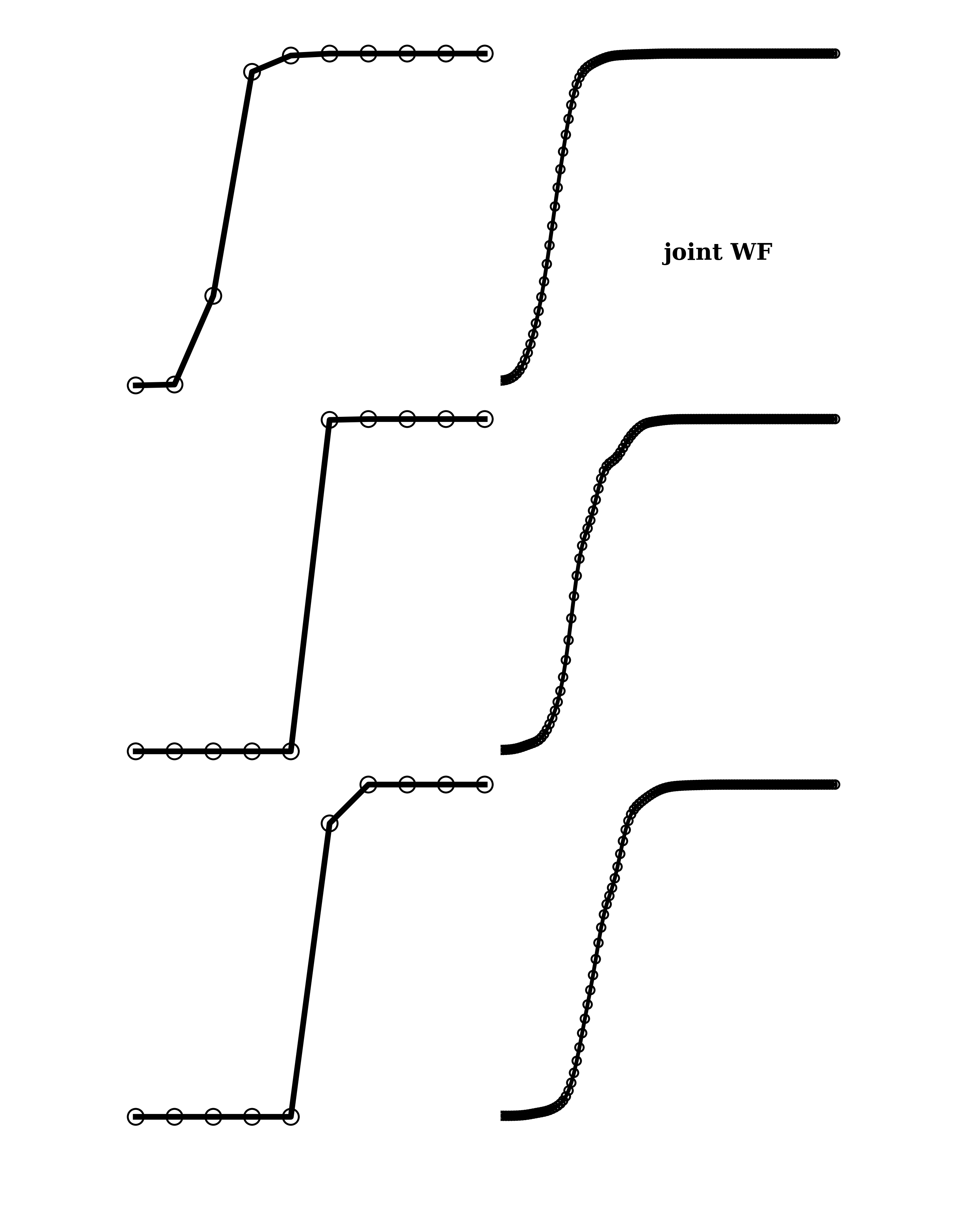}%
  \end{ocg}
  \hspace{-0.65\hsize}%
  \caption{\emph{Left column}: the probability of the injected periodic signal at $P_c=8.962$ days existing 
    in synthetic RV time-series---as a function of the injected semi-amplitude $K_c$---with time-sampling
    identical to the published \ToggleLayer{fig:Hon,fig:Hoff}{\protect\cdbox{HARPS WF}} \citepalias{cloutier17b},
    the published \ToggleLayer{fig:Con,fig:Coff}{\protect\cdbox{CARMENES WF}} \citepalias{sarkis18}, or their
    \ToggleLayer{fig:jon,fig:joff}{\protect\cdbox{joint WF}}. Three sets of synthetic
    RV time-series are considered and contain K2-18c only (\emph{top row}), K2-18b and c (\emph{middle row}),
    or both planets plus a GP correlated noise model of stellar activity (\emph{bottom row}). The
    \emph{shaded vertical region} highlights the MAP and $1\sigma$ measured value of $K_c=4.63 \pm 0.72$ \mps{}
    from \citetalias{cloutier17b}.
    \emph{Right column}: the probability of the injected periodic signal at $P_c$ existing
    in synthetic RV time-series---with fixed $K_c=4.63$ \mps{---}as a function of the number of RV measurements.}
  \label{fig:9vKcNrv}
\end{figure*}

For each synthetic RV time-series we compute the Bayesian generalized Lomb-Scargle periodogram
\citep[GLSP;][]{mortier15} from which we isolate the probability of a sinusoidal function with the period of
K2-18c ($P_c=8.962$ days) being present in our synthetic time-series; p($P_c|$RV).
The left column of Fig.~\ref{fig:9vKcNrv} depicts p($P_c|$RV) as a function of the injected K2-18c semi-amplitude for
three out of the four RV models. 
The synthetic time-series containing K2-18b and stellar activity are not included in Fig.~\ref{fig:9vKcNrv} as
they were consistently seen to result in p($P_c|$RV)$\ll 1$\% thus indicating that the $P_c$ did not arise with any
significance when not explicitly added to the time-series. The ordinate values in
Fig.~\ref{fig:9vKcNrv} are the median probabilities derived from a set of 50 synthetic time-series realizations
per value of the injected $K_c$. In this way, we marginalize over the exact form of the injected white and correlated
noise sources which are sampled randomly in each of the 50 iterations.
As expected, because the $P_c$ periodic signal is injected into each synthetic time-series,
the probability of that signal existing within the data increases with the $K_c$ from effectively zero probability
when $K_c\sim 1$ \mps{} towards p($P_c|$RV) $= 100$\% as $K_c \to 10$ \mps{} for any of the three types of synthetic
time-series. It is true that as the complexity of the synthetic time-series
increases (i.e. as more signals are added) the semi-amplitude $K_c$ needs to be larger in order to be
detected with high probability. It is also clear that detecting the injected $P_c$ signal is easier with either
the HARPS or joint WFs as their probability curves tend to increase more rapidly with $K_c$
and they approach 100\% probability at a lower $K_c$ than with the CARMENES WF alone. This is
particularly true at the MAP value of $K_c=4.63$ \mps{} \citepalias{cloutier17b} wherein
p($P_c|$RV) is $\sim 40$\% larger with the HARPS WF than with CARMENES for any of the synthetic
time-series. This shows that with the CARMENES time-sampling 
the strength of the $P_c$ periodic signal is less prominent in the GLSP than with the HARPS---or
joint---time-sampling. With any of the three types of synthetic time-series, the strength
of $P_c$ is typically lower with CARMENES until $K_c\sim 10$ \mps{} wherein the probability
of $P_c$ with CARMENES becomes consistent with 100\%. However, an injected value of $K_c=10$ \mps{} is
inconsistent with the \citetalias{cloutier17b} measured value at $\gtrsim 7\sigma$.

The systematically lower $P_c$ probability with CARMENES may be due to sampling, instrumental effects, or to the fact
that the CARMENES WF contains fewer RVs; 58 compared to 75
with HARPS. The smaller WF affects the sampling of periodic signals and $P_c$ may not be strongly
detectable with only 58 RVs.
To investigate this possibility, we again compute p($P_c|$RV) in our synthetic RV time-series but for
random subsets of each time-series and with an increasing number of RV measurements 
$N_{\text{RV}} \in [10,N_f]$ where $N_f$ is the full size
of each RV time-series.\footnote{i.e. $N_f =$ 75, 58, and 133 for HARPS, CARMENES, and their joint time-series
  respectively.} When creating these synthetic times-series, $K_c$ is fixed to its MAP value of 4.63 \mps{.}
The smoothed probability curves for each synthetic time-series and each WF are shown in the right
column of Fig.~\ref{fig:9vKcNrv}. The curves are smoothed to remove the high frequency noise and make the trends in the
curves easier to parse visually. As can be seen in the probability of $P_c$ as a function of $K_c$, when $K_c$ equals
its MAP value, the $P_c$ signal is detected at a higher probability with the HARPS or HARPS+CARMENES
WFs than with CARMENES alone. Here we focus on the probability of $P_c$ when the HARPS and CARMENES time-series
contain the same number of measurements.
When both time-series are equal to the size of the full CARMENES WF (i.e. $N_{\text{RV}}=58$), the
probability of detecting $P_c$ is always smallest with the CARMENES WF than with any subset of 58 measurements with
either the HARPS or joint WFs. For the most realistic set of synthetic RVs featuring two planets + a stellar activity
model, the discrepancy in p($P_c$|RV) is modest with HARPS being $\sim 29$\% greater than with CARMENES and
their joint WF being $\sim 63$\% greater.

Overall we see that the probability of the $P_c$ periodic signal existing in time-series with the sampling of HARPS,
CARMENES, or their joint time-series, is systematically lowest with CARMENES. By the nature of this experiment we
conclude that the sole reason for the lower CARMENES probability is due to its WF. Although this discrepancy hints
at why  $P_c$ may not have been detected in the GLSP of the CARMENES RVs, the relative values of p($P_c$|RV) to
surrounding periodicities in these synthetic RVs is considered high and is certainly sufficient to detect $P_c$.
However, next we show that a small subset of anomalous CARMENES observations are likely responsible for the suppression
of the $P_c$ periodic signal in the GLSP.

\subsection{Identifying anomalous CARMENES observations} \label{sect:anomalous}
\begin{figure*}
  \centering
  \includegraphics[width=\hsize]{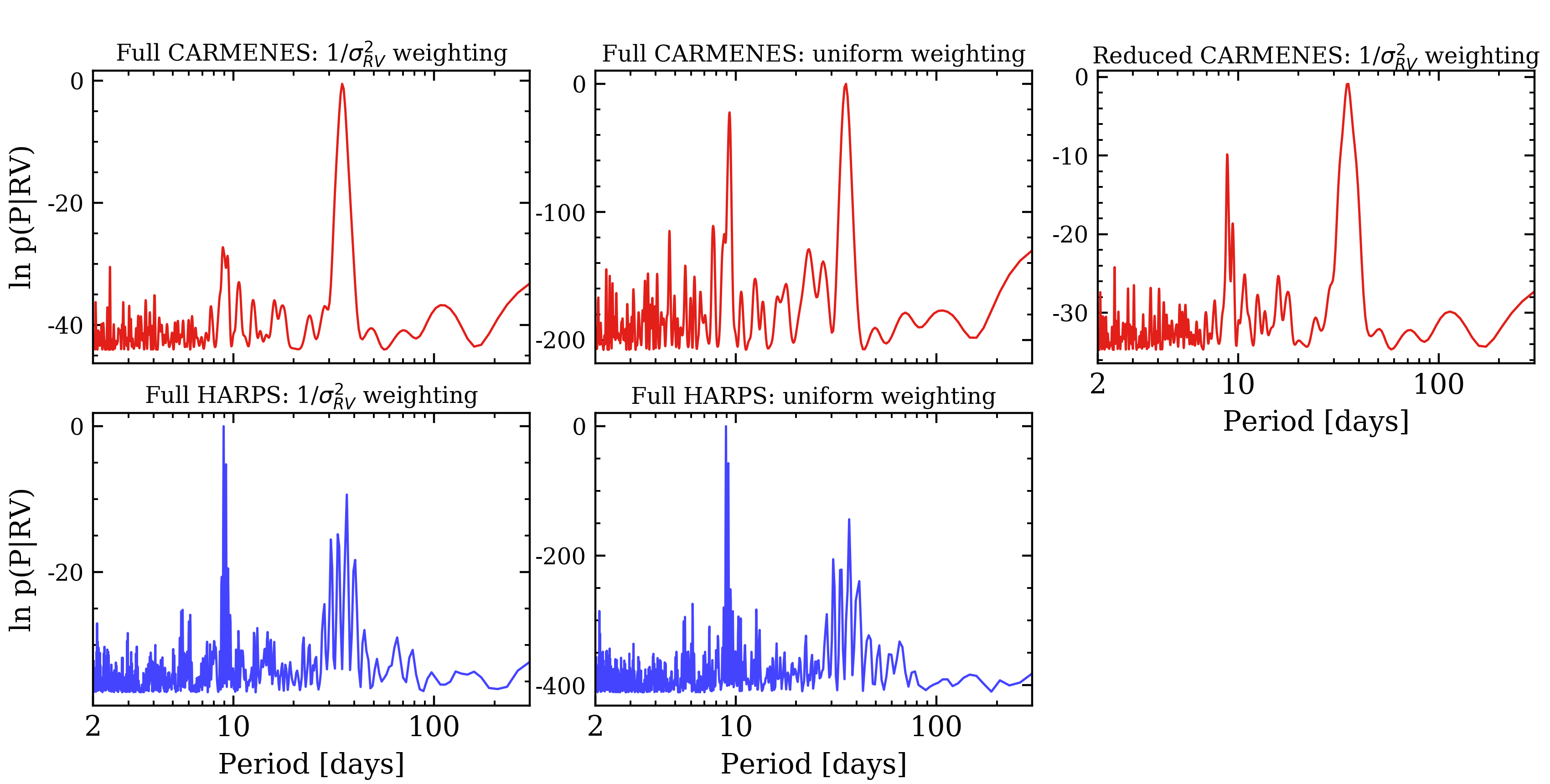}%
  \hspace{-\hsize}%
  \begin{ocg}{fig:linesoff}{fig:linesoff}{0}%
  \end{ocg}%
  \begin{ocg}{fig:lineson}{fig:lineson}{1}%
  \includegraphics[width=\hsize]{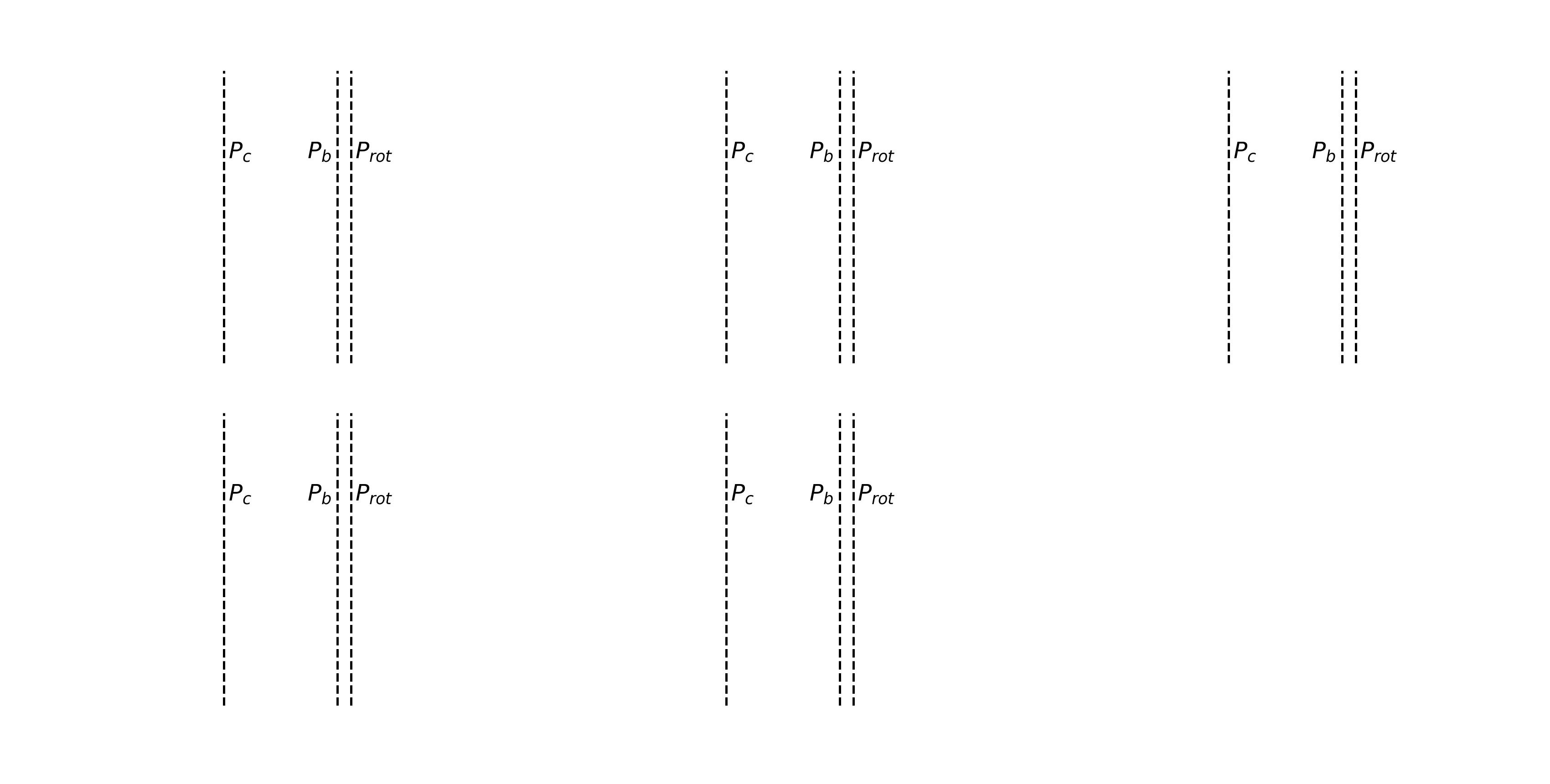}%
  \end{ocg}
  \hspace{-\hsize}%
  \caption{Bayesian generalized Lomb-Scargle periodograms for various subsets of the published CARMENES and HARPS RVs 
    with one of a pair of possible weighting schemes. The details of the time-series shown in each panel are
    annotated above the panel. The three
    \ToggleLayer{fig:lineson,fig:linesoff}{\protect\cdbox{\emph{dashed vertical lines}}} depict the orbital
    period of the proposed non-transiting planet K2-18c ($P_c=8.962$ days), the orbital period of the known
    transiting planet K2-18b ($P_b=32.93963$ days), and the photometric stellar rotation period
    ($P_{\text{rot}}=38.6$ days). The $P_c$ signal posited to be due to a second, non-transiting planet
    is seen at high relative probability in all but the full CARMENES RV time-series from \citetalias{sarkis18}
    with a $1/\sigma_{\text{RV}}^2$ weighting.}
  \label{fig:glsp}
\end{figure*}

Recall that the periodic signal from the proposed planet K2-18c at $P_c=8.962$ days was not seen with a low
false alarm probability in the GLSP of the full CARMENES time-series \citepalias{sarkis18}. This is confirmed in the
first panel of Fig.~\ref{fig:glsp} although a small (albeit non-significant) hint of the $\sim 9$ day signal is visible.
In computing the GLSP the CARMENES RVs are weighted by the inverse square of their respective measurement
uncertainties. As a brief experiment, we considered what the effect of adopting a uniform weighting on each RV (i.e.
unweighted) would have on the probability of
the 9 day signal. As can be seen in the second panel of Fig.~\ref{fig:glsp}, the 9 day signal becomes much more significant
when using a uniform weighting. For comparison, the probability of the 9 day signal in the HARPS GLSP varies only weakly
between the weighted and unweighted conventions (c.f. bottom row of Fig.~\ref{fig:glsp}).
This suggests that the 9 day signal \emph{does} exist within the CARMENES RV
dataset despite only appearing with significance when using an unconventional---and incorrect---method of computing the
GLSP.

\begin{figure}
  \centering
  \includegraphics[width=\hsize]{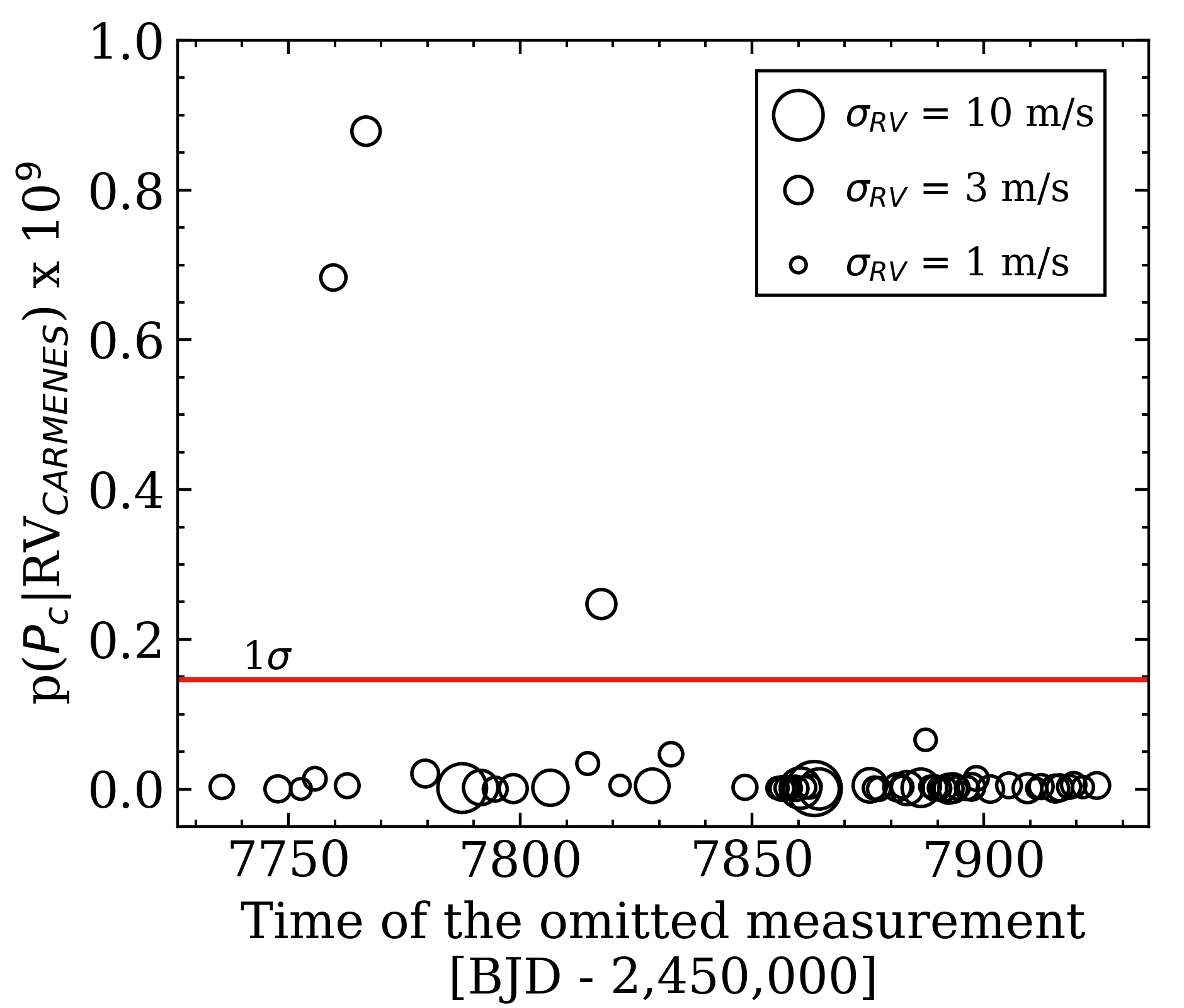}
  \caption{The probability of the proposed K2-18c periodic signal $P_c=8.962$ days existing within the
    CARMENES RV dataset from \citetalias{sarkis18} but with a single measurement omitted via leave-one-out
    cross-validation. The abscissa depicts
    the observation epoch of each omitted RV measurement. The \emph{solid horizontal line} depicts the $1\sigma$
    dispersion of the probabilities. The three measurements which lie above the $1\sigma$ line significantly
    suppress the probability of $P_c$ and are henceforth treated as anomalous.}
  \label{fig:prob9}
\end{figure}

The sudden appearance of the 9 day periodic signal in the CARMENES RV suggests that some anomalous 
measurements may be partially responsible for the signal's suppression
to the extent that it becomes buried in the noise of the full CARMENES GLSP. If the number of such
anomalous measurements is small compared to the full size of the dataset, then we can justify the removal those
measurements to measure the 9 day signal with CARMENES given our strong prior evidence for the signal from HARPS
\citepalias{cloutier17b}.
We proceed by calculating the probability of $P_c$ existing within various subsets of the full CARMENES time-series
via leave-one-out cross-validation.
In each of the 58 considered subsets, we omit a single unique measurement, compute the
GLSP of the remaining 57 RVs, and isolate the probability of $P_c$ existing within the data using an identical method
to what was used in Sect.~\ref{sect:sampling}. The resulting probabilities of $P_c$ as a function of the epoch of the
omitted measurement are shown in Fig.~\ref{fig:prob9}. 

In Fig.~\ref{fig:prob9}
we identify three anomalous RVs via a visual $\sigma$-clip\footnote{For RV indices starting at 0, the three anomalous
  CARMENES RVs have indices 4, 6, and 14 (i.e. BJD-2,450,000 = 7759.69656, 7766.73773, 7817.51320).}.
We note that we refer to these measurements as anomalous as their inclusion
versus their omittance clearly results in a significant reduction in p($P_c|$RV) which is not seen for the majority of
the CARMENES RVs. These measurements have associated RV uncertainties that are
comparable to the mean CARMENES RV measurement uncertainty and thus have a significant effect on the probabilities
of the periodicities sampled in the GLSP. The removal of these three anomalous measurements and the
recalculation of the GLSP---using the proper RV weighting---is shown in the third panel of Fig.~\ref{fig:glsp}.
The 9 day periodic signal is now clearly seen at high probability. Clearly the strategic removal of just
3 out of 58 CARMENES RVs enhances the $P_c$ periodic signal. Thus we have significant preliminary evidence for the
existence of the proposed planet K2-18c at $\sim 9$ days from the GLSP of the remaining 55 CARMENES RVs.

\citetalias{sarkis18} provided their contemporaneous spectroscopic time-series of the CARMENES `full', 
blue, and red RVs, as well as time-series of the chromospheric  H$\alpha$ index and the three Ca$_{\text{II}}$ infrared
triplet line indices. Inspection of these time-series does not reveal any obvious reason for why the three
measurements identified in Fig.~\ref{fig:prob9} significantly suppress the 9 day signal.
We shared this result among the CARMENES team
members who were also unable to identify any potential causes of the anomalous nature of these measurements
after inspecting the measured RVs in individual orders. Therefore, at this time we are unable to explain
the cause of the anomalous nature of these three measurements.

A similar exercise as shown in Fig.~\ref{fig:prob9} was also conducted using the HARPS RVs. The results of which
are not presented here because the removal of individual HARPS RVs did not result
in any significant changes to the probability of $P_c$ existing within the reduced dataset; i.e. all values of
p($P_c$|RV$_{\text{HARPS}}$) were close to 100\% with a small rms of $\sim 6$\%.
The discrepancy between HARPS and CARMENES in this regard may be because the 9 day signal
is less suppressed by the HARPS WF compared to the CARMENES WF (c.f. Fig.~\ref{fig:9vKcNrv}) or because the HARPS
WF contains more measurements and is thus less sensitive to the removal of individual measurements. The latter
scenario highlights the need to obtain large $N_{\text{RV}}$ when searching for small planets whose RV
semi-amplitudes are comparable to the RV measurement precision. This result has also been noted in simulations of
`blind' RV searches \citep[e.g.][]{cloutier18} that strongly advocate for `more RVs per star' rather than `more stars
with fewer RVs per star' in order to maximize future discoveries of small RV planets.

\section{Chromatic dependence of the 9 day signal with HARPS} \label{sect:act}
In addition to the RV variations derived from the 42 CARMENES visible orders, \citetalias{sarkis18} also derived RVs from
the first and second halves of these orders spanning 561-689 and 697-905 nm respectively. Signal variations between
these \emph{blue} and \emph{red} RVs may elude to the nature of those signals as stellar activity or from achromatic
dynamical influences from planetary companions.
Fluctuations in the strength of the 9 day signal in the CARMENES RVs helped lead \citetalias{sarkis18}
to conclude that the signal is due to stellar activity because of its apparent wavelength dependence. However, this
evidence does not rule out the possibility that instead the 9 day signal is planetary in nature and appears to vary
between the blue and red RVs because its suppression by activity is chromatically variable.

Similarly to \citetalias{sarkis18}, here we compute the chromatic HARPS RVs to investigate the dependence of the
9 day signal strength with wavelength. The method used to derive these RVs at each observation epoch is detailed
in Sect. 2.1 of \citetalias{cloutier17b} and is based on the methodology from \cite{astudillodefru15}. 
The HARPS RVs are re-derived in each of the 72 HARPS orders although we restrict our analysis to
orders redder than 498 nm where the signal-to-noise ratio (S/N) per spectral order is sufficient to reach a
$\sigma_{\text{RV}}$ per order $\lesssim 30$ \mps{.} The RVs derived from the remaining 34 orders are then grouped into
blue and red orders whose weighted mean is used to compute the blue and red HARPS RVs. Our chromatic HARPS RVs
span uneven wavelength ranges of 498-594 nm and 618-688 nm such that the resulting median RV measurement precision of
$\sim 7$ \mps{} is comparable between the two sets of RVs. Note that the wavelength domain spanned by the red HARPS RVs
is approximately equal to the redder half of the blue CARMENES wavelength domain.

\begin{figure}
  \centering
  \includegraphics[width=\hsize]{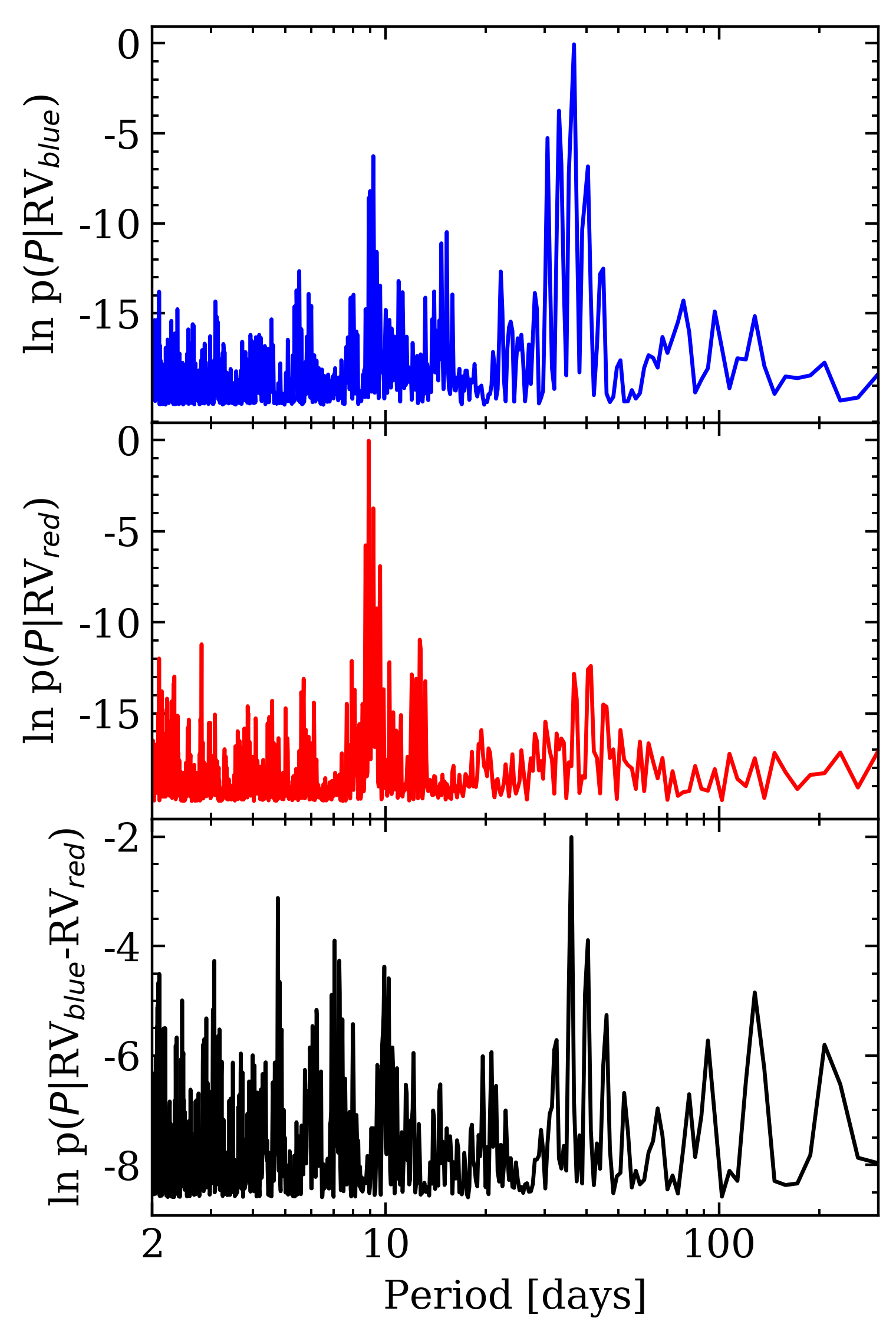}%
  \hspace{-\hsize}%
  \begin{ocg}{fig:linesoff}{fig:linesoff}{0}%
  \end{ocg}%
  \begin{ocg}{fig:lineson}{fig:lineson}{1}%
  \includegraphics[width=\hsize]{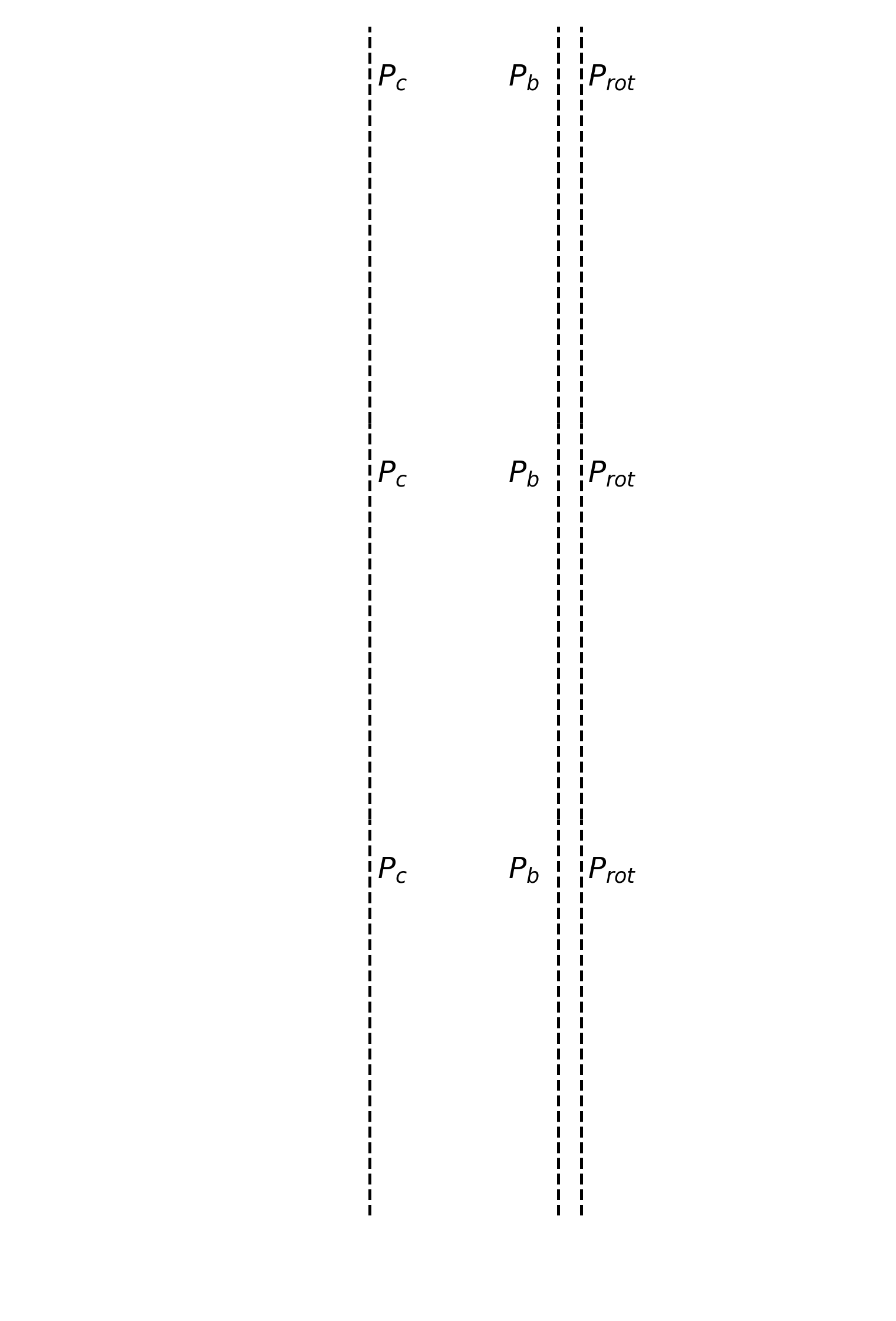}%
  \end{ocg}
  \hspace{-\hsize}%
  \caption{Bayesian generalized Lomb-Scargle periodograms of the blue (\emph{top}), red (\emph{middle}), 
    HARPS RVs, and their difference (\emph{bottom}). The three
    \ToggleLayer{fig:lineson,fig:linesoff}{\protect\cdbox{\emph{dashed vertical lines}}} depict the orbital
    period of the planets K2-18b and c ($P_c\sim 9$ days), and the photometric stellar rotation period.
    The 9 day signal is seen in the first two time-series but at a lower probability in the blue likely due to the
    higher levels of stellar activity in that wavelength regime.
    The 9 day signal is suppressed in the GLSP of the RV difference while some residual probability close to \prot{}
    continues to persist due to the incomplete removal of stellar activity.}
  \label{fig:glspchrom}
\end{figure}

The GLSPs of the blue and red HARPS RVs are shown in Fig.~\ref{fig:glspchrom}.
In both GLSPs the $\sim 9$ day signal is discernible along with the forest of peaks around $P_b$ and the stellar
rotation period due to aliasing from the HARPS WF \citepalias[c.f. Fig.2][]{cloutier17b}.
Most notably, the probability of the 9 day peak is significantly greater in the HARPS red RVs compared to in the blue.
This is expected if the 9 day signal is indeed due to a
planet---whose signal strength is achromatic---whereas stellar activity arising from the temperature contrast of active
regions is expected to increase bluewards \citep{reiners10} thus degrading the S/N of the planetary signal in the blue
RVs relative to the red. As such, if the 9 day signal was originating from stellar activity rather than from a
planet, one would expect the 9 day periodic signal to be stronger in the blue RVs which it is not.
Indeed the rms of the blue RVs is slightly greater than in the red (7.8 \mps{} compared to 6.9 \mps{})
despite each set of chromatic RVs having comparable S/N. We note that this excess dispersion in the blue HARPS RVs is
only marginal given the star's moderate activity level \citepalias[$\sim 2.7$ \mps{)}][]{cloutier17b,sarkis18}
which is less than RV measurement precision in either the blue or red HARPS RVs ($\sim 7$ \mps{)}.
The stronger activity level seen in the blue is likely responsible for the decreased significance
of the 9 day signal and the enhanced probability at the stellar rotation period compared to the red.

Furthermore in Fig.~\ref{fig:glspchrom}, we include the GLSP of the blue minus red RVs. The 
9 day signal is significantly suppressed whereas some residual probability close to the stellar rotation period persists
along with some residual probability near $P_b$ due to the aliasing of $P_{\text{rot}}$ by the HARPS WF.
The suppression of the 9 day signal in the differential RVs is indicative of its achromatic nature
(i.e. a dynamical signal) whereas the differing signal strength of RV activity in the blue and red RVs results in some
residual power close to \prot{.} This further supports the planetary interpretation of the 9 day signal. 

\section{Temporal dependence of the 9 day signal with HARPS} \label{sect:temporal}
In addition to the proposed chromatic dependence of the 9 day signal, \citetalias{sarkis18} addressed
the possibility that the 9 day signal strength also varies with time. This was posited based on the increased 
strength of the 9 day peak in the GLSP of the second half of the CARMENES RVs compared to the first. However as was
shown in Sect.~\ref{sect:anomalous}, three anomalous CARMENES RVs exist in the first half of the CARMENES WF
that significantly suppress the 9 day signal in the GLSP. This naturally explains why a stark increase in the 9 day
signal strength was seen in the latter half of the CARMENES WF rather than being due to temporal variability in the
stellar activity.

To further investigate the dependence of the 9 day signal on activity with HARPS, we can consider HARPS activity
indices and the probability of the 9 day signal in each HARPS observing season separately.
To extend the investigation of the temporal dependence of the 9 day signal we obtained 31 additional HARPS
spectra of K2-18 (i.e. in addition to the 75 presented in \citetalias{cloutier17b}).
These new spectra extend the full HARPS baseline from April 2015 (BJD=2457117.5) to July 2018
(BJD=2458307.5). The method used to derive the stellar RVs at each observation epoch is detailed
in Sect. 2.1 of \citetalias{cloutier17b}. The full set of 106 HARPS RVs are provided in Table~\ref{table:data}.

\begin{figure*}
  \centering
  \includegraphics[width=\hsize]{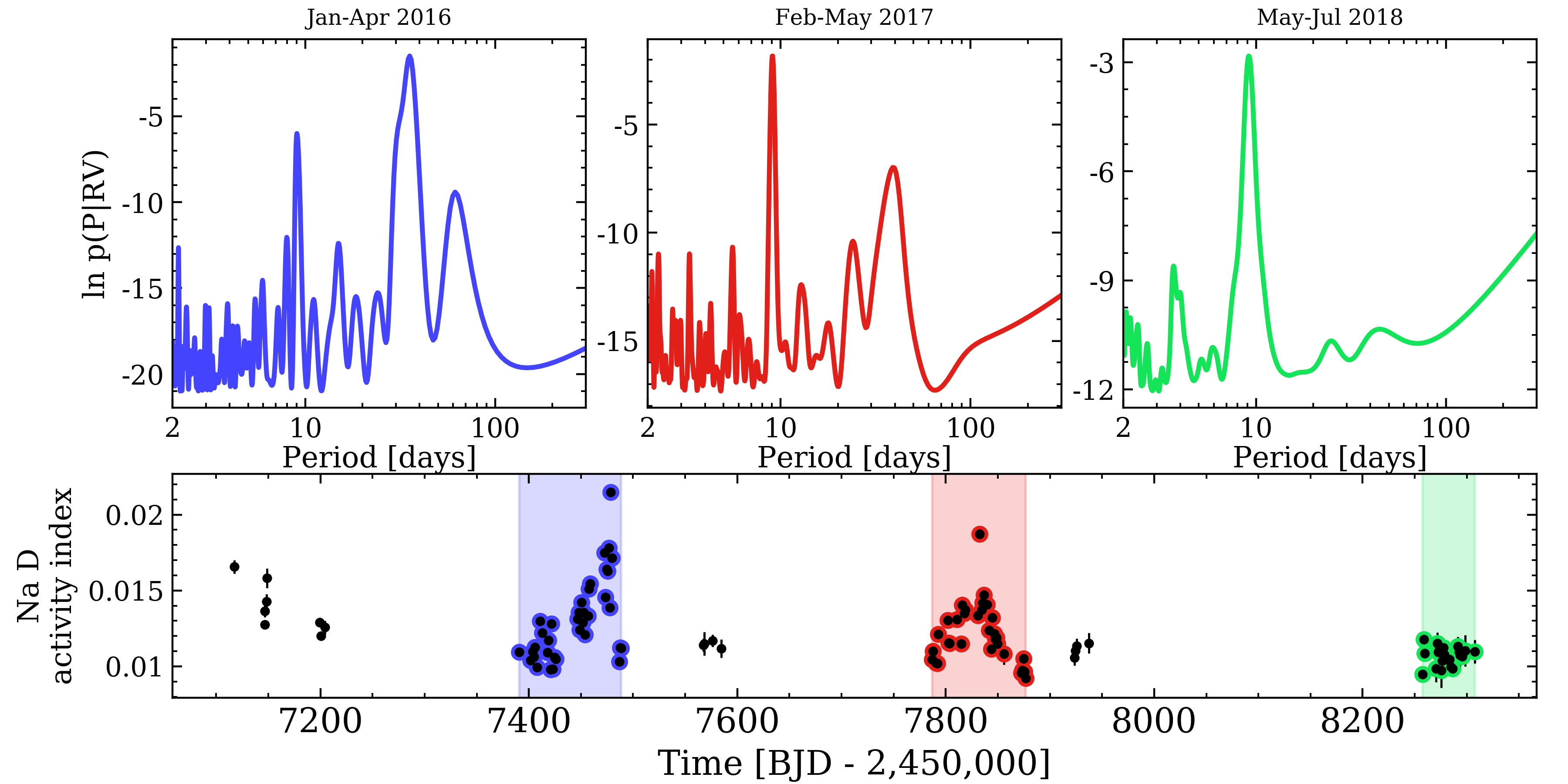}%
  \hspace{-\hsize}%
  \begin{ocg}{fig:vertoff}{fig:vertoff}{0}%
  \end{ocg}%
  \begin{ocg}{fig:verton}{fig:verton}{1}%
  \includegraphics[width=\hsize]{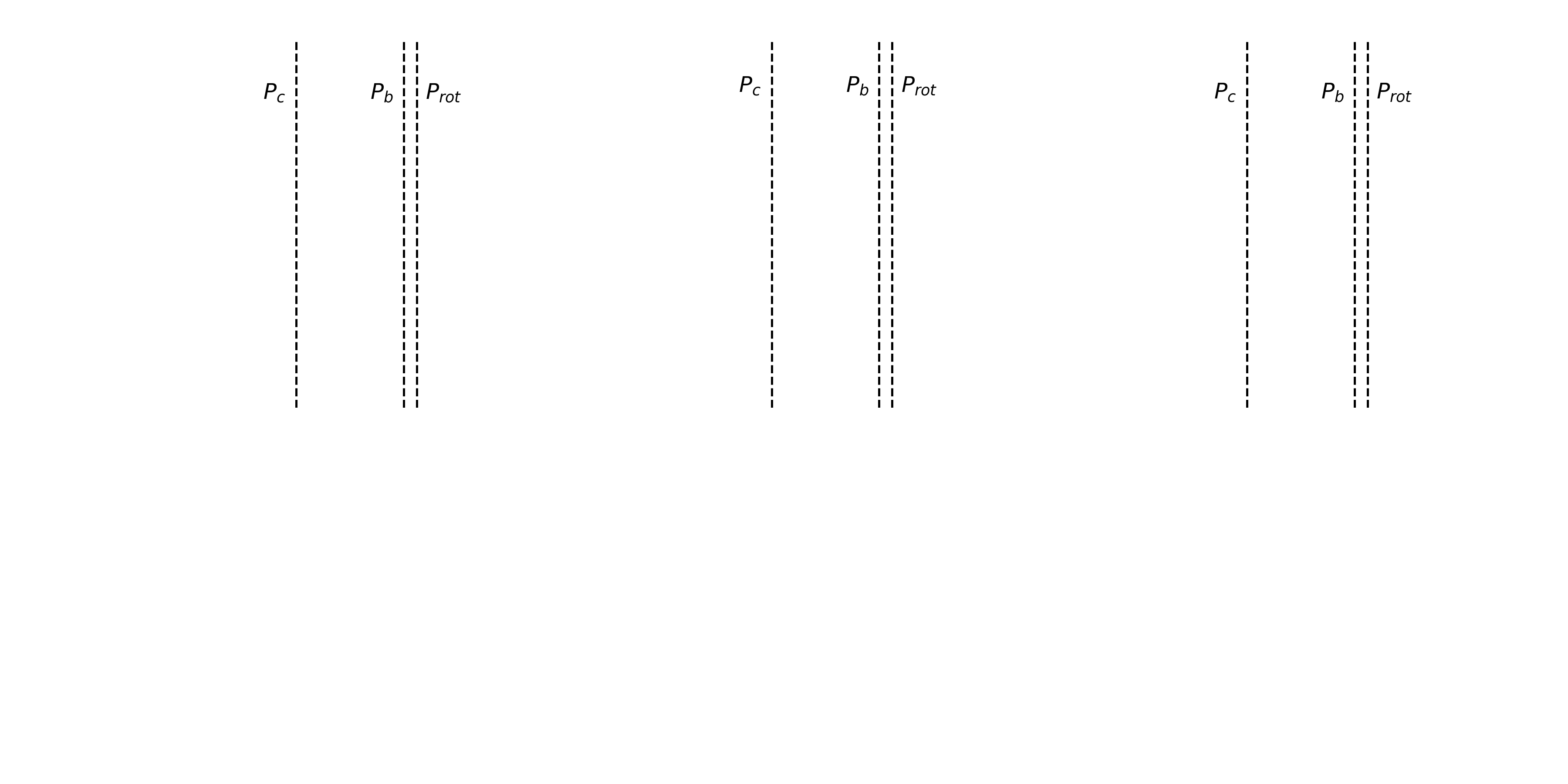}%
  \end{ocg}
  \hspace{-\hsize}%
  \begin{ocg}{fig:textoff}{fig:textoff}{0}%
  \end{ocg}%
  \begin{ocg}{fig:texton}{fig:texton}{1}%
  \includegraphics[width=\hsize]{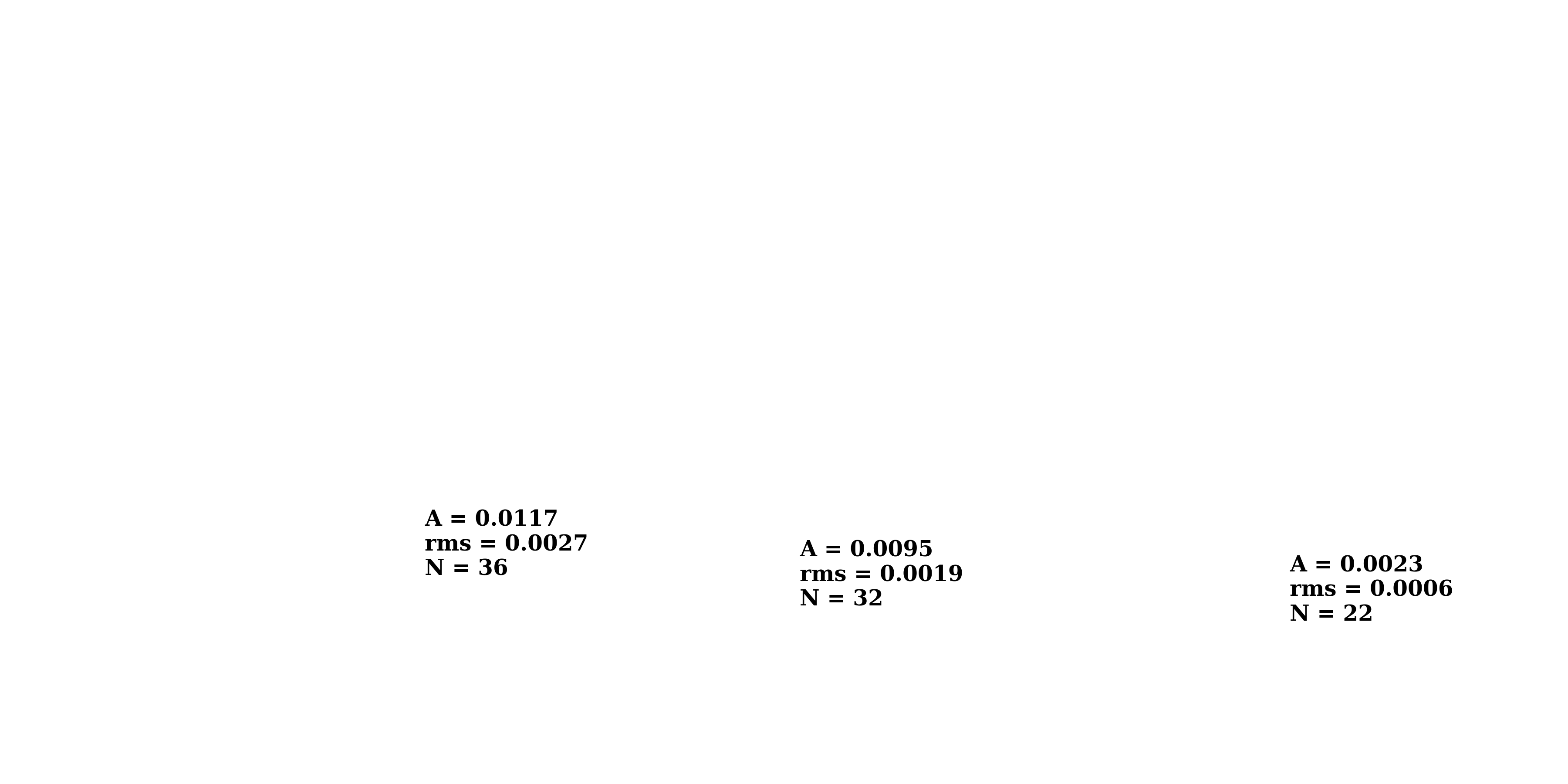}%
  \end{ocg}
  \hspace{-\hsize}%
  \caption{\emph{Top row}: Bayesian generalized Lomb-Scargle periodograms of the HARPS RVs in the three observing
    seasons annotated above each panel. The
    \ToggleLayer{fig:verton,fig:vertoff}{\protect\cdbox{\emph{vertical dashed lines}}} depict
    the orbital period of the proposed non-transiting planet K2-18c ($P_c = 8.962$ days), the orbital period
    of the known transiting planet K2-18b ($P_b = 32.93963$ days), and the photometric stellar rotation period (\prot{}
    $= 38.6$ days). \emph{Bottom row}: the sodium doublet time-series as measured by HARPS. The coloured regions/markers
    are indicative of the epochs used to compute each RV GLSP in the upper row. The 
    \ToggleLayer{fig:texton,fig:textoff}{\protect\cdbox{\emph{annotation group}}} adjacent to each observing season
    depicts the Na D peak-to-peak amplitude $A$, the Na D rms, and the number of measurements within that observing season.
    This $A$ and rms diagnostics indicate that the level of stellar activity is decreasing with time while the $P_c$ signal
    is simultaneously becoming more prominent.}
  \label{fig:temporal}
\end{figure*}

The full HARPS time-series is spanned by three separate observing seasons
containing $N_{\text{RV}} \geq 22$. The GLSPs of the HARPS RVs in each observing season are shown in
Fig.~\ref{fig:temporal}. Although the 9 day signal is visible in each GLSP, its probability relative to the
surrounding continuum is seen to increase with time from early 2016 to mid-2018.
If the 9 day signal is planetary in nature rather than being
due to stellar activity, then we would expect the K2-18 activity level to decrease with time thus enhancing the 9
day signal in the GLSP as the activity level subsides. Next we will show that this is indeed the case.

To characterize the temporal variability of the K2-18 activity level we compute the strength of the sodium doublet
activity index (Na D) in all HARPS spectra following \cite{astudillodefru17b}. The Na D time-series is shown in the lower
panel of Fig.~\ref{fig:temporal}. In particular we focus on the peak-to-peak amplitude $A$ and rms of the Na D measurements in each
observing season. In doing so we see that the amplitude of
the variation in the Na D activity index and its rms both decrease across subsequent observing seasons. Specifically, we find that
$A=0.0117$ in the first observing season and drops to 0.0023 after $\sim 26$ months. Similarly, the Na D rms drops from 0.0027
to 0.0006 over the same time interval. These diagnostics indicate that indeed
the level of stellar activity is decreasing with time and thus supports the planetary interpretation of the 9 day signal.
A similar trend of increasing activity is also observed when considering other activity
indicators such as the H$\alpha$ index although its time-series is not depicted in Fig.~\ref{fig:temporal}.

\section{Simultaneous RV modelling of planets and correlated `noise'} \label{sect:correlated}
In the era of ultra-precise RV spectrographs whose inherent stability often operates below the photon-noise limit,
RV detections of small planets such as K2-18c are limited by nuisance signals from stellar activity. Numerous
techniques have been tested to mitigate the effects of stellar activity whose amplitude and quasi-periodic temporal
variability can mask and/or mimic planetary signals. Such techniques include linear correlations with
contemporaneous activity indicators \citep[e.g.][]{boisse09},
pre-whitening \citep[e.g.][]{queloz09},
parametric modelling of stellar surface features \citep[e.g.][]{dumusque14},
and sine wave fitting such as that used in \citetalias{sarkis18}. The main issue with the latter technique is that
the rotationally modulated activity in photometry and in RVs
is not strictly periodic as the finite lifetimes of active regions, along with
their variable sizes, contrasts, and spatial distributions will introduce a quasi-periodic component. This is
especially true when RV time-series span many stellar rotation cycles.
Incomplete models can result in the miscalculation of planetary parameters and the marginalization of coherent
signals (e.g. additional planets) that are required to properly interpret the observed RV variations. When modelling RVs
it is therefore crucial to include a flexible model that can account for stochastic variations in stellar
activity. This is effectively done in a non-parametric way using Gaussian process (GP) regression simultaneously with
planetary models (i.e. keplerians) thus ensuring self-consistent solutions between planets and stellar activity.
Furthermore, GP modelling fits within a Bayesian formalism as a single GP---describing the temporal covariance between RV
measurements with a single set of hyperparameters---is itself a prior distribution of functions whose mean represents
the `best-fit' activity model \citep{haywood14,faria16,cloutier17a}.
Here we analyze a variety of RV time-series from either the HARPS
\citepalias{cloutier17b} or CARMENES \citepalias{sarkis18} spectrographs using a model that includes one or two
planets plus a correlated `noise' component from stellar activity in the form of a GP regression model.

Our full 2-planet model with observations taken by a single spectrograph contains 16 model parameters: the systemic
velocity $\gamma$, an additive scalar jitter $s$, four quasi-periodic GP hyperparameters
$\{a, \lambda, \Gamma, P_{\text{GP}}\}$, and five keplerian parameters per planet
$\{P, T_0, K, h=\sqrt{e}\cos{\omega}, k=\sqrt{e}\sin{\omega} \}$. For cases in which we combine observations from
HARPS and CARMENES we treat their activity models as separate GPs \citep[e.g.][]{grunblatt15}
owing to their unique systematics, the chromatic dependence of stellar activity, and each spectrograph's distinct
wavelength coverage. In this case,
all GP hyperparameters are common between the two GP models with the exception of the additive jitter and the covariance
amplitude. When modelling the joint HARPS+CARMENES time-series we therefore have 19 model parameters.

The GP regression models of stellar activity are trained on the star's precision K2 photometry. The
apparent photometric variability---from which the photometric stellar rotation period was measured
($P_{\text{RV}}=38.6$ days; \citetalias{cloutier17b})---is sensitive
to photospheric active regions which also have an observable manifestation in the RVs with common covariance
properties. However, we note that photometry is only weakly sensitive to chromospheric plages which also
contribute to RV activity signals, at least in Sun-like stars \citep{haywood16}. We
use the K2 photometry to train our GP stellar activity models to ensure that the
mean GP model from the simultaneous planet + activity modelling is representative of stellar
activity and does not settle into a solution that describes other temporally correlated signals (e.g.
non-transiting planets) by restricting the $P_{\text{GP}}$ to \prot{} or one of its low-order
harmonics. By training our GP on ancillary time-series we empirically constrain the covariance structure of the
activity signal and use the posterior probability density functions (PDFs)
of the GP hyperparameters from training as priors during
the RV modelling stage (see Table~\ref{table:priors}).

\begin{table}
\small
\renewcommand{\arraystretch}{0.7}
\centering
\caption[]{Summary of the RV model parameter priors used for all models throughout this study}
\label{table:priors}
\begin{tabular}{lc}
\hline \\ [-1ex]
Parameter & Prior \smallskip \\
\hline \\ [-1ex]
Systemic velocity, $\gamma$ [m/s] & $\mathcal{U}(\bar{\mathbf{RV}}-10, \bar{\mathbf{RV}}+10)$ \smallskip \\
\emph{GP hyperparameters} & \\
Covariance amplitude, & \\
$\ln{(a/}$(m/s)) & $\mathcal{U}(-3,3)$ \\
Exponential timescale, & \\
$\ln{(\lambda/}$days) & $\text{p}(\ln{\lambda}|\text{K2 photometry})$ \\
Coherence, $\ln{(\Gamma)}$ & $\text{p}(\ln{\Gamma}|\text{K2 photometry})$ \\
Periodic timescale, & \\
$\ln{(P_{\text{GP}}/}$days) & $\text{p}(\ln{P_{\text{GP}}}|\text{K2 photometry})$  \\
Additive jitter, $s$ [\mps{]} & $\mathcal{U}(0,10)$ \smallskip \\
\emph{Keplerian parameters} & \\
$P_b$ [days] & $\mathcal{N}(32.93961,10^{-4})^{(\bullet)}$ \\
$T_{0,b}$ [BJD-2,450,000] & $\mathcal{N}(7264.3914,6.3 \times 10^{-4})^{(\bullet)}$ \\
$K_b$ [\mps{]} & $\text{\emph{mod}}\mathcal{J}(1,20)^{(\ast)}$ \\
$h_b = \sqrt{e_b}\cos{\omega_b}$ & $\mathcal{U}(-1,1)^{(\dagger)}$ \\
$k_b = \sqrt{e_b}\sin{\omega_b}$ & $\mathcal{U}(-1,1)^{(\dagger)}$ \\
$P_c$ [days] & $\mathcal{U}(8,10)$ \\
$T_{0,c}$ [BJD-2,450,000] & $\mathcal{U}(7259,7269)$ \\
$K_c$ [\mps{]} & $\text{\emph{mod}}\mathcal{J}(1,10)^{(\ast)}$ \\
$h_c = \sqrt{e_c}\cos{\omega_c}$ & $\mathcal{U}(-1,1)^{(\dagger)}$ \\
$k_c = \sqrt{e_c}\sin{\omega_c}$ & $\mathcal{U}(-1,1)^{(\dagger)}$ \medskip \\
\hline
\end{tabular}
\begin{list}{}{}
\item {\bf{Notes.}} 
  $^{(\bullet)}$ based on the transit light curve measurements from \cite{benneke17}. \\
  $^{(\ast)}$ $\text{\emph{mod}}\mathcal{J}(k,l)$ \mps{}
  refers to a modified Jeffreys prior on a parameter $A$ which behaves like a uniform
  prior for $A \ll$ the knee at k \mps{} and
  behaves like a Jeffreys prior at $A \gg k$ up to $l$. We use a modified Jeffreys prior on
  the RV semi-amplitudes $K$ to sample multiple decades as a Jeffreys prior but also include
  $K=0$ \mps{} which a Jeffreys prior does not \citep{gregory05}. \\
  $^{(\dagger)}$ We further insist
  that $e = h^2 + k^2 < 1$.
\end{list}
\end{table}

In these analyzes we sample the posterior PDFs
of the RV model parameters given an input dataset via Markov chain Monte-Carlo (MCMC) simulations.
All simulations are run using the affine-invariant MCMC ensemble sampler
\texttt{emcee} \citep{foremanmackey13}. All model parameters are initialized around their MAP values
with $1\sigma$ dispersions from \citetalias{cloutier17b}. The adopted model parameters are consistent
between the various time-series considered and are summarized in Table~\ref{table:priors}. In each
MCMC simulation we manually monitor the acceptance fraction and ensure that it always lies between
20-50\% for both the burn-in phase and throughout the actual posterior PDF sampling.

\subsection{CARMENES RVs} \label{sect:analysisC}
Here we model the subset of the CARMENES-visible RVs presented in \citetalias{sarkis18} which are known to
not result in the anomalous suppression of the 9 day signal. We consider two RV models, each containing
a quasi-periodic GP regression model of stellar activity. The first model contains only one planetary
signal from the confirmed transiting planet K2-18b while the second model additionally includes the
second planet K2-18c at $\sim 9$ days.
The RVs and GLSPs are plotted in Fig.~\ref{fig:analysisC} for both the one and two planet models after
iteratively removing the MAP models of activity and planetary signals.

In the 1-planet model of the 55 CARMENES RVs, the GP activity model has a covariance amplitude of
7.5 \mps{} which is greater than the sinusoidal amplitude of 2.7 \mps{} measured by \citetalias{sarkis18}
on nearly the same dataset. Based on the GLSP of K2-18b (i.e. with activity removed), it is clear that
although the activity model has a large amplitude, it fails to model the 9 day signal. The
GLSP of the residuals following the removal of activity and K2-18b ($K_b=3.61 \pm 0.82$ \mps{)} clearly
exhibits a strong periodic signal at $\sim 9$ days hinting at the existence of an additional signal that
is unmodelled when assuming a 1-planet model.

The stellar activity in the 2-planet model has a similarly large covariance amplitude of 8.2 \mps{.}
However the only significant signal in the GLSP of the RV activity is at the stellar rotation period.
Similarly the GLSP of K2-18b ($K_b=2.91\pm 0.88$ \mps{)}
only exhibits a significant signal at $P_b$ and the GLSP of K2-18c ($K_c=2.31\pm 0.76$ \mps{)} exhibit
a strong signal at $\sim 9$ days with a somewhat weaker signal at $\sim 5.5$ days. Indeed the GLSP of the
residuals following the removal of both planets and activity only shows a significant residual probability at
$\sim 5.5$ days which only arises after the removal of activity and K2-18b (c.f. panels of O-C and K2-18c in
Fig.~\ref{fig:analysisC}). The nature of this signal is less obvious as---unlike the 9 day signal---it does
not appear with enough significance in either GLSP of the HARPS or CARMENES RVs prior to the removal of any
modelled signals (c.f. Fig.~\ref{fig:glsp}). One possible explanation is that the $\sim 5.5$ day
signal arises from an alias of $P_c$ with the CARMENES WF which exhibits excess power close to the baseline
duration of $\sim 189$ days. Using the standard formula to compute the alias frequency from the signal and WF
frequencies (i.e. $f_{\text{alias}} = f_{\text{signal}} + n f_{\text{WF}}$),
and setting $f_{\text{signal}}=1/8.997$ days$^{-1}$ and
$f_{\text{WF}}=1/189$ days$^{-1}$, we find an aliased periodicity at $\sim 5.56$ days when $n=13$.
Given the high-order $n$ required to identify an aliased periodicity that is seemingly consistent with the excess
probability at $\sim 5.5$ days, we do not claim that this WF alias explains the signal's origin and similarly
we cannot discard the possibility that the 5.5 day signal comes from an additional planet that is insofar undetected.
More RV data are required to investigate the source of this signal. In Sect.~\ref{sect:evidence}
we will perform a model comparison considering the possibility that the 5.5 day signal is due to a third planet
in the system.

\subsection{All HARPS RVs} \label{sect:newrvs}
In Sect.~\ref{sect:temporal} we presented 31 new HARPS RVs to investigate the temporal variability of the
9 day signal. Hence the full HARPS WF has been extended to over a year past the previously most recent
published HARPS measurement for this system \citepalias{cloutier17b} and now contains 106 RV measurements.
Here we model the full HARPS time-series identically to as was done for the CARMENES RVs in
Sect.~\ref{sect:analysisC}. The RVs and GLSPs are plotted in Fig.~\ref{fig:analysisH106}.

In the 1-planet model the GP activity model has a covariance amplitude of 2.3 \mps{,} comparable to the
MAP $K_b=2.75\pm 0.66$ \mps{.}
Similarly to the 1-planet model of the CARMENES RVs, the activity model fails to account for
the high probability of the 9 day signal. The 9 day peak continues to persist following the removal of
the K2-18b keplerian.  

In the 2-planet model the GP activity model has a somewhat larger covariance amplitude compared to
the 1-planet model; 4.18 \mps{.} This amplitude is comparable to the MAP semi-amplitudes of the
the two planets ($K_b=3.32\pm 0.60$ \mps{,} $K_c=3.71\pm 0.57$ \mps{)} and,
given the proximity of the stellar rotation period to $P_b$ and aliases of the two aforementioned periods with
the WF \citepalias{sarkis18}, the activity model only partially suppresses the GLSP probabilities between
$\sim 30-50$ days. It is also clear that when the mean activity model and only a single planet are
removed, the only remaining signal at high probability is that of the remaining planet at 9 days. Furthermore, it
is clear that there are no residual signals at high probability when all modelled signals are removed. 
Most notably, a probability peak at $\sim 5.5$ days---as was seen in the CARMENES residuals with a
2-planet model (Fig.~\ref{fig:analysisC})---is visible but only at the level of the noise.

\subsection{Joint HARPS+CARMENES RVs}
Here we model the joint RV time-series of the 106 HARPS plus the 55 CARMENES RVs.
The RVs and GLSPs are plotted in Fig.~\ref{fig:analysisHC}. In the 1-planet model the
covariance amplitude of the HARPS and CARMENES stellar activity models are 1.5 and 5.5 \mps{} respectively.
These values are each slightly smaller than the covariance amplitudes measured when considering each
spectrograph's time-series individually but their ratio is nearly preserved.
Similarly to either spectrograph's individual RV analysis in the presence of a 1-planet
model, the GLSP of the residuals following the removal of K2-18b ($K_b=3.00\pm 0.50$ \mps{)}
and activity exhibits a strong periodic signal
at $\sim 9$ days which again hints at the existence of an additional planetary signal.

The stellar activity covariance amplitudes in the 2-planet model are comparable to as in the 1-planet model; i.e.
3.0 and 5.5 \mps{} for HARPS and CARMENES respectively.
The corresponding GLSP of the RV activity is reminiscent of the 1-planet RV activity GLSP with the exception
that the inclusion of two modelled planets ($K_b=2.75\pm 0.43$ \mps{,} $K_c=2.76\pm 0.41$ \mps{)} 
drastically reduces the probability of the 9 day signal.
Indeed in the GLSP of K2-18c, the strongest signal is at $\sim 9$ days with only a hint of the $\sim 5.5$
day signal that was seen in CARMENES.
In both the GLSP of the HARPS and joint RV residuals following the removal of both planets and
activity (c.f. Figs.~\ref{fig:analysisH106} and~\ref{fig:analysisHC}),
the $\sim 5.5$ day signal is not seen at high probability which suggests that the signal is not physical
and instead arises stochastically as a by-product of the CARMENES WF.

\subsection{Overlapping HARPS \& CARMENES window functions}
For a maximally one-to-one comparison we can compare the RV model analyzes and GLSP structures
in the subsets of the HARPS and CARMENES RVs that are restricted to the 138 days from February 2nd to
June 20th, 2017. Between these dates the HARPS and CARMENES WFs overlap such that we
have approximately contemporaneous RVs taken with each spectrograph. By only considering
the observations taken throughout the overlapping time span we minimize our sensitivity to temporal
variations in stellar activity whose properties may vary between successive observing cycles. The overlapping
WF contains 35 HARPS and 50 CARMENES RVs. One of the CARMENES RVs in the overlapping window
was found to anomalously suppress the $\sim 9$ day signal in Sect.~\ref{sect:anomalous} so we discard it and
are left with 49 CARMENES RVs. The RVs and GLSPs are plotted in Fig.~\ref{fig:analysisHCoverlap}.

In the 1-planet model the covariance amplitudes are equivalent with
each spectrograph (i.e. 2.0 \mps{)} and are notably small compared to the previously
analyzed time-series. This may be due to the lack of a long-term near-linear trend in the stellar activity
over the short time span considered here.
The corresponding activity model appears close to flat indicating that the RV activity
has only weak structure over this relatively short time span. The low activity amplitude also results in
a low probability at $P_{\text{rot}}$ and the activity GLSP being dominated by the 9 day signal which is effectively
unmodelled when only 1 planet is considered.
We measure $K_b=3.96\pm 0.73$ \mps{} which along with the activity model reveals the residual 9 day signal
as well as the $\sim 5.5$ day signal that was seen in the CARMENES residuals.

In the 2-planet model the covariance amplitudes are nearly identical to the 1-planet model (i.e. 2.0 \mps{)}
and therefore exhibit a similarly featureless structure. The small covariance amplitudes of the activity
models result in the activity GLSP containing primarily noise. Comparatively, the GLSPs of the modelled planets
($K_b=3.59\pm 0.62$ \mps{,} $K_c=2.65\pm 0.58$ \mps{)} are dominated
by their respective periodicities with the $\sim 5.5$ days signal appearing in the GLSP
of K2-18c, albeit at a much lower probability than the 9 day signal. However in the residual GLSP, the
$\sim 5.5$ day signal is largely suppressed after removing K2-18c. 

\subsection{Model comparison} \label{sect:evidence}
The detection of exoplanets in RV data is fundamentally based on whether or not the input dataset favours the
existence of the planet of interest. This is typically done within a Bayesian framework wherein the fully marginalized
likelihoods (i.e. the evidence) of competing models (i.e. 1 versus 2 planets) are computed and used for model comparison.
In this formalism, a planet is said to be `detected' if the evidence for the $n+1$ planet model is significantly larger
than the evidence for a model containing $n$ planets. Here we calculate the model evidences for the purpose of model
comparison and use the resulting values to determine whether or not the putative RV planet K2-18c is favoured by
the various time-series considered.

Each model's Bayesian evidence is approximated using the estimator from \cite{perrakis13} and the marginalized
posterior PDFs from our MCMC analyses as importance samplers. The \cite{perrakis13} estimator is known to result in
quantitatively similar results to other more robust but computationally expensive methods
\citep[e.g. nested samplers;][]{nelson18}. Model comparison requires that all common model parameters
between competing models be drawn from identical prior distributions which are explicitly reported in
Table~\ref{table:priors}. Our Bayesian evidence estimates are reported in Table~\ref{table:perrakis} for both the 1
and 2-planet models and for all input time-series considered.

Also included in Table~\ref{table:priors} are the 2-1 Bayes factors (i.e. evidence ratios) of the
2-planet model relative to 1-planet to infer if the second planet K2-18c is favoured or disfavoured by the
corresponding time-series. Overall, we find that the explicit values of the Bayesian
evidence favour the 2-planet model for \emph{all} time-series considered. However, the dispersion in calculated
evidence values when using various methods of calculation are known to vary by factors of $\gtrsim 10^2$
depending on the complexity of the model (i.e. the number of planets; \citealt{nelson18}). Recall that the
simplest model considered in this study is not the 0-planet model as we know from the transit light curves
that K2-18b exists at $\sim 33$ days. Effectively, we are therefore only tasked with detecting one new RV planet
rather than two. But given the caveat that uncertainties in the calculated evidence
can be of order $10^2$, we require that the evidence ratio of the 2-planet model to the 1-planet model must
be $\geq 10^2$ for the second planet K2-18c to be `detected'. Under this condition there are two instances
in which K2-18c is not detected. The first occurs with the full set of the 58 CARMENES RVs from
\citetalias{sarkis18} in which K2-18c is not detected due to the three anomalous measurements identified
in Sect.~\ref{sect:anomalous}. This result is consistent with the null detection of K2-18c with these data
in \citetalias{sarkis18}. However, the 2-1 Bayes factor for CARMENES alone exceeds $10^2$ following the
removal of the three aforementioned measurements. Secondly, the blue CARMENES RVs only weakly favour a
second planet which can be attributed to the increased RV rms at these shorter wavelengths\footnote{i.e.
  7.5 \mps{} compared to 5.14 and 5.73 \mps{} in the full and red CARMENES RVs respectively.}. This trend
is seen again in the blue and red HARPS RVs for which a second planet is more strongly favoured
by the red RVs where the RV rms is smaller. The increased measurement uncertainty for CARMENES in the
blue hides planetary signals and makes the inference of their presence less certain given
the correspondingly low data likelihoods.

\begin{table*}
\small
\renewcommand{\arraystretch}{0.7}
\centering
\caption[]{Marginal likelihood estimations and Bayes factors for various RV datasets and models}
\label{table:perrakis}
\begin{tabular}{ccccc}
\hline \\ [-1ex]
Dataset & $N_{\text{RV}}$ & Model & $\ln$ Model evidence$^{(\bullet)}$ & Bayes factor: 2 to 1 planets$^{(\ast)}$ \\
$D$ & & $\mathcal{M}_i$ & $\ln{\text{p}(D|\mathcal{M}_i)}$ & $\text{p}(\mathcal{M}_2|D)/ \text{p}(\mathcal{M}_1|D)$ \smallskip \\
\hline \\ [-1ex]
HARPS & 106 & 1 planet + GP & -338.5 & - \\
- & - & 2 planets + GP & -325.5 & $3\times 10^5$ \\
CARMENES & 58 & 1 planet + GP & -180.6 & - \\
- & - & 2 planets + GP & -178.1 & 7 \\
reduced CARMENES & 55 & 1 planet + GP & -169.5 & - \\
- & - & 2 planets + GP & -164.1 & 143 \\
HARPS + reduced CARMENES & 161 & 1 planet + GP & -489.6 & - \\
- & - & 2 planets + GP & -475.8 & $6\times 10^5$ \\
\hline \\ [-1ex]
blue HARPS & 106 & 1 planet + GP & -375.5 & - \\
- & - & 2 planets + GP & -369.2 & 336 \\
red HARPS & 106 & 1 planet + GP & -375.5 & - \\
- & - & 2 planets + GP & -359.4 & $6\times 10^6$ \\
blue CARMENES & 55 & 1 planet + GP & -186.8 & - \\
- & - & 2 planets + GP & -183.0 & 28 \\
red CARMENES & 55 & 1 planet + GP & -174.0 & - \\
- & - & 2 planets + GP & -159.1 & $2\times 10^6$ \\
\hline \\ [-1ex]
HARPS (Feb-June 2017) & 35 & 1 planet + GP & -121.0 & - \\
- & - & 2 planets + GP & -113.6 & 1018 \\
CARMENES (Feb-June 2017) & 49 & 1 planet + GP & -151.7 & - \\
- & - & 2 planets + GP & -146.0 & 181 \\
HARPS + CARMENES (Feb-June 2017) & 84 & 1 planet + GP & -256.5 & - \\
- & - & 2 planets + GP & -249.6 & 613 \\
\hline
\end{tabular}
\begin{list}{}{}
\item {\bf{Notes.}}
      $^{(\bullet)}$ Estimates of the model evidences are calculated using the \cite{perrakis13} estimator and the
  marginalized posterior probability density functions from our MCMC runs. \\
  $^{(\ast)}$ Bayes factors---or evidence ratios---are written as
  $\frac{\text{p}(\mathcal{M}_2|D)}{\text{p}(\mathcal{M}_1|D)} = \frac{\text{p}(D|\mathcal{M}_2)}{\text{p}(D|\mathcal{M}_1)} \frac{\text{p}(\mathcal{M}_2)}{\text{p}(\mathcal{M}_1)}$ where each model prior is $\text{p}(\mathcal{M}_i) = \alpha^i$ for
  $\alpha = (\sqrt{5}-1)/2 \approx 0.618$ such that $\sum_{i=1}^{2} \text{p}(\mathcal{M}_i) = 1$.
\end{list}
\end{table*}

Recall the $\sim 5.5$ day signal seen in the K2-18c and residual GLSPs of the RV time-series containing
CARMENES data in Figs.~\ref{fig:analysisC}~\ref{fig:analysisHC}, and~\ref{fig:analysisHCoverlap}. 
As a test of the potential planetary origin of this signal we first ran an MCMC on the CARMENES RVs as it is
there that the residual 5.5 day signal exhibited the highest probability in the GLSP following the removal of K2-18b,
c, and stellar activity. Similarly to the 1 and 2-planet models
we then estimate the evidence of this 3-planet model using the estimator from \cite{perrakis13} and compare it to the
2-planet model for the same input time-series. For the third planet
we adopt identical priors to that of K2-18c (c.f. Table~\ref{table:priors})
with the exception of the planet's orbital period and time of mid-conjunction which are modified to
$\mathcal{U}(4.5,6.5)$ days and $\mathcal{U}(7259,7265.5)$ BJD-2,450,000 respectively.
The resulting $\ln$ evidence for the 2 and 3-planet models are
-165.0 and -162.4 respectively. The corresponding 3-2 Bayes factor is $\sim 7$ implying that the 3-planet model
including a planet at $\sim 5.5$ days is not significantly favoured over the 2-planet model. By a similar exercise
using the full joint HARPS+CARMENES time-series yields a 3-2 Bayes factor of $\sim 0.8$. Therefore by the effective
accounting of the 5.5 day periodic signal by our K2-18c models in Figs.~\ref{fig:analysisH106},~\ref{fig:analysisHC},
and~\ref{fig:analysisHCoverlap}, and the disfavourability of the 3-planet model compared to just 2 planets, we conclude
that a third planet at $\sim 5.5$ days is not detected in the available RV data but whose signal origin may be alluded to
with additional RV monitoring.

\section{Discussion and conclusions} \label{sect:dis}
We have conducted a systematic re-analysis of the published HARPS \citepalias{cloutier17b} and CARMENES
\citepalias{sarkis18} RVs of the transiting planet host K2-18
to identify the source of the apparent 9 day signal which---prior to this study---was
only seen in the former dataset. We have also included an additional set of 31 new HARPS RVs to
investigate the temporal dependence of the 9 day signal and improve the measurement precision of
planet parameters. The following are our main conclusions:

\begin{enumerate}
\item The CARMENES window function is somewhat detrimental to the detection of an injected 9 day keplerian
  signal---compared to the HARPS window function---in that the injected signal is seen at a lower probability in
  the generalized Lomb-Scargle periodogram (GLSP) when using the CARMENES window function.
\item The cause of the non-detection of the 9 day signal in \citetalias{sarkis18} was shown to result from
  three anomalous CARMENES measurements, the removal of which reveals the existence of the 9 day signal in the GLSP
  of the remaining 55 RVs.
\item We computed two sets chromatic HARPS RVs. The 9 day
  signal is seen in both time-series and at a significantly higher probability in the red HARPS RVs where stellar
  activity is weaker. This supports the planetary interpretation of the 9 day signal.
\item The 9 day signal is retrieved with HARPS in each of its three observing seasons separated by $\sim 1$ year.
  The probability of the 9 day signal increases with time simultaneously with a decrease in the level of stellar
  activity as probed by the Na D activity index. This further supports the planetary interpretation of the 9 day
  signal.
\item We adopt a non-parametric stellar activity model to account for stellar variability over the multiple stellar
  rotation cycles spanned by the observations, and simultaneously model activity and planetary signals. This results in
  self-consistent planet solutions and the ability to compare 1 and 2-planet models on equal grounds.
\item In all considered times-series, the Bayesian model evidence favours a 2-planet model over the 1-planet model which
  includes K2-18c at $\sim 9$ days.
\end{enumerate}

By the points listed above, we have obtained compelling evidence for the planetary nature of the 9 day signal seen in HARPS and
in the reduced CARMENES RV time-series. It is important to highlight the importance of basing RV planet
detections off of robust Bayesian model comparison tests rather than basing those detections
solely off of periodogram false alarm probabilities (FAPs) which can vary stochastically and are highly sensitive to variations
in the input time-series (e.g. weighting schemes). Although significant peaks in a GLSP are useful for the initial
identification of periodic signals in unevenly sampled time-series, conclusions regarding their actual existence and origin
should not be made solely based on their FAP. Accurate and simultaneous modelling of all signals present in a
time-series is required to infer accurate model parameters of planets and activity. Furthermore,
Bayes factors---or the ratio of competing models' fully
marginalized likelihoods---are robust model comparison tools which marginalize over all prior information about
models with competing numbers of planets and penalize overly complicated models. In this way they are optimally
suited to the detection confirmation of periodic planetary signals.

\begin{figure}
  \centering
  \includegraphics[width=\hsize]{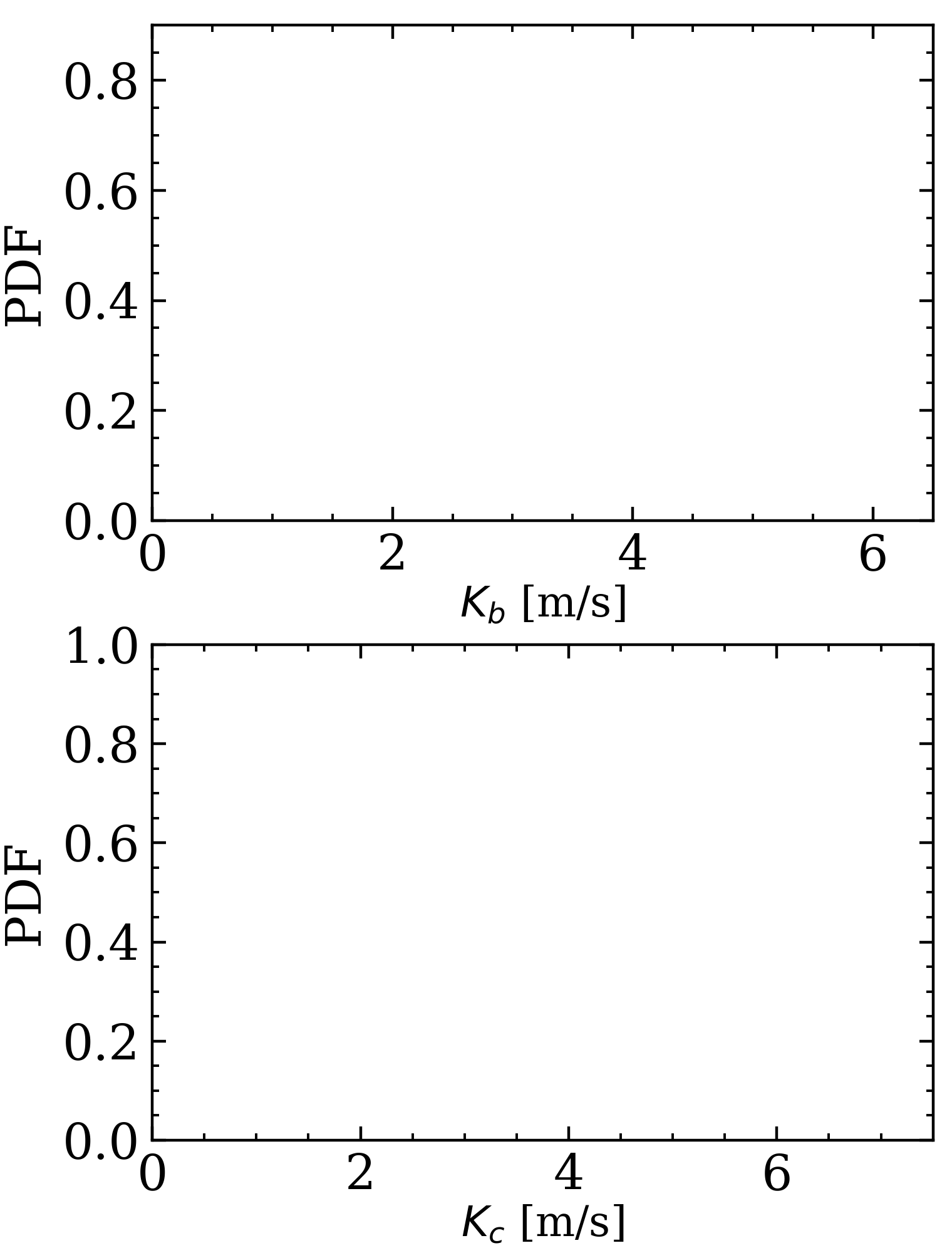}%
  \hspace{-\hsize}%
  \begin{ocg}{fig:HKoff}{fig:HKoff}{0}%
  \end{ocg}%
  \begin{ocg}{fig:HKon}{fig:HKon}{1}%
  \includegraphics[width=\hsize]{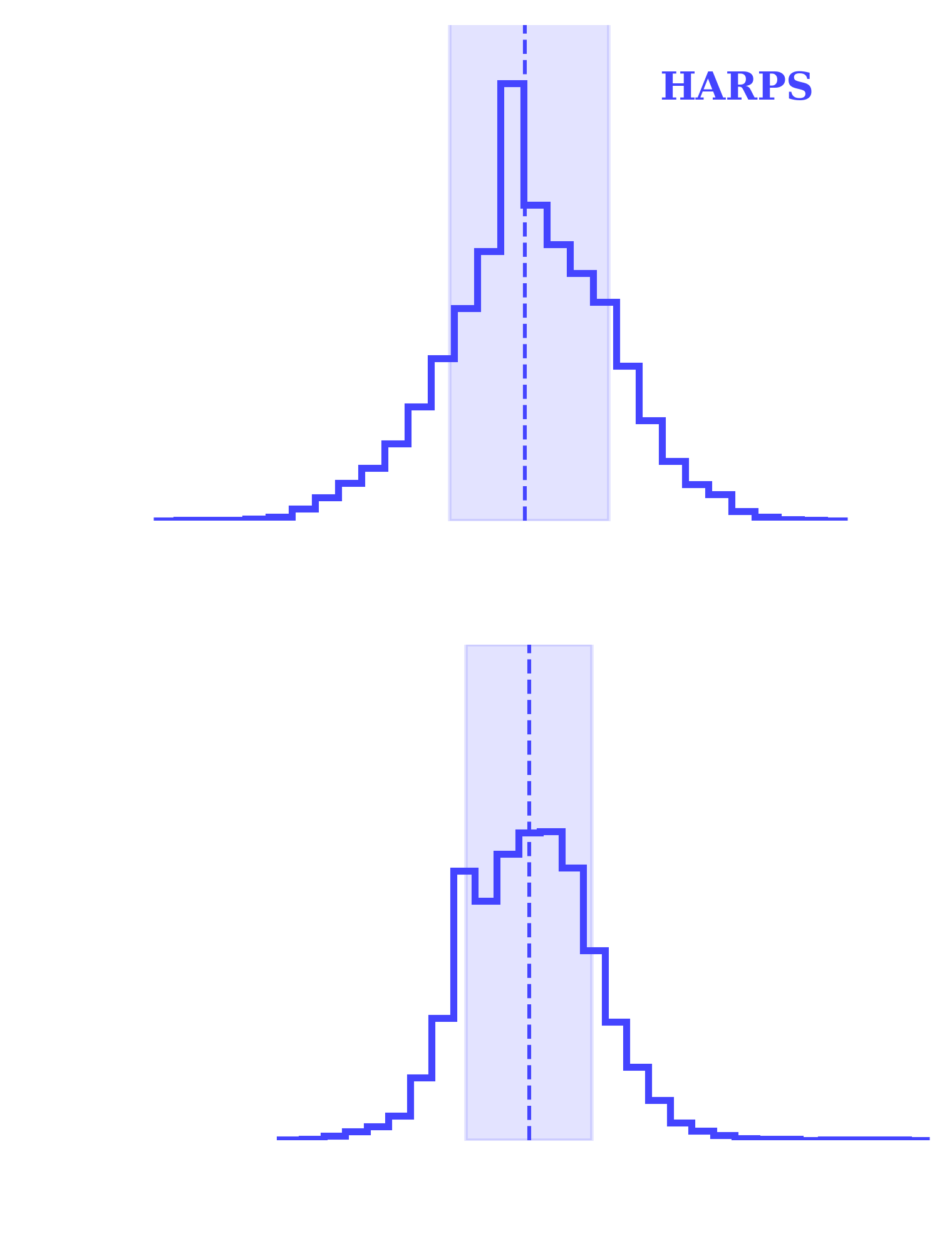}%
  \end{ocg}
  \hspace{-\hsize}%
  \begin{ocg}{fig:CKoff}{fig:CKoff}{0}%
  \end{ocg}%
  \begin{ocg}{fig:CKon}{fig:CKon}{1}%
  \includegraphics[width=\hsize]{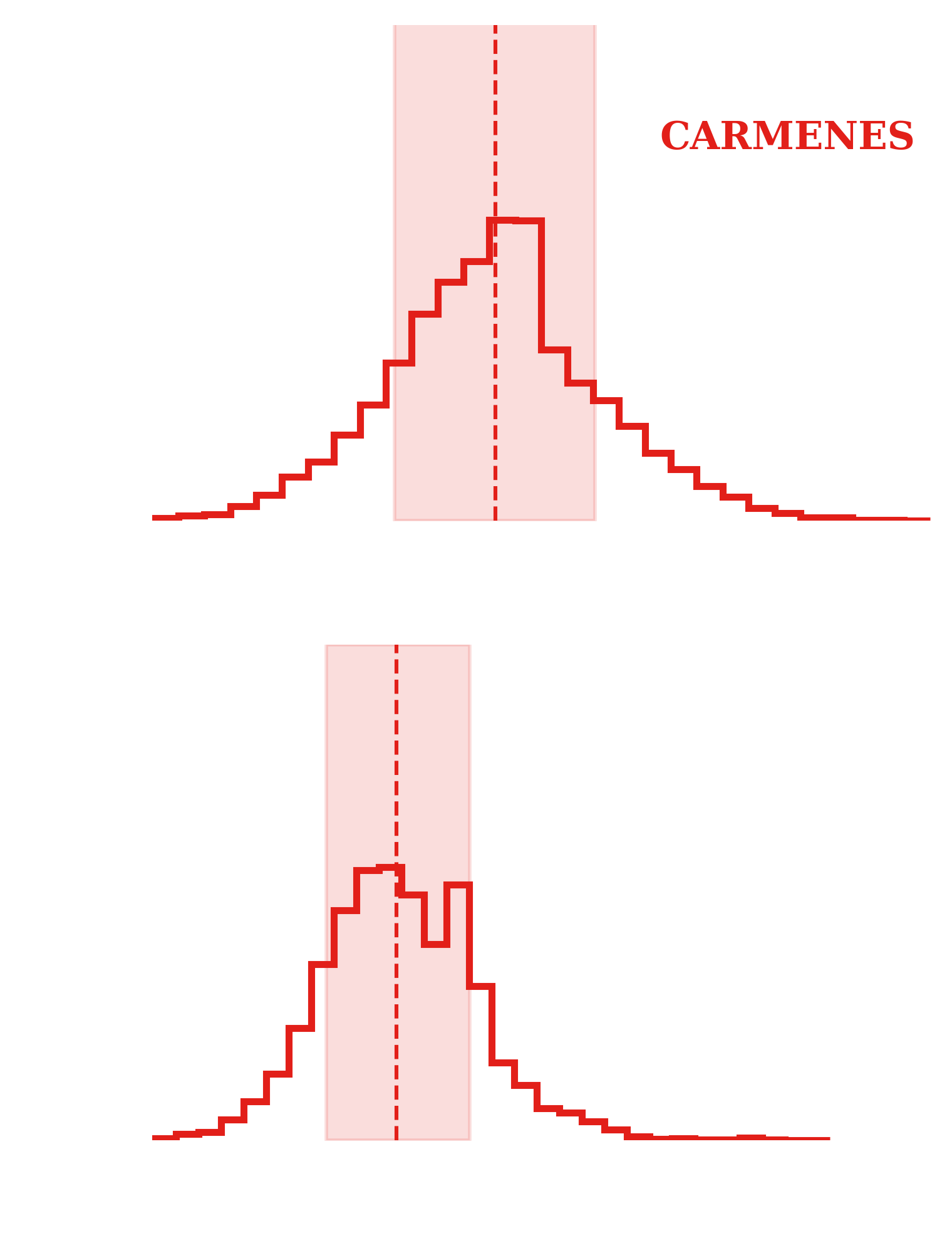}%
  \end{ocg}
  \hspace{-\hsize}%
  \begin{ocg}{fig:jKoff}{fig:jKoff}{0}%
  \end{ocg}%
  \begin{ocg}{fig:jKon}{fig:jKon}{1}%
  \includegraphics[width=\hsize]{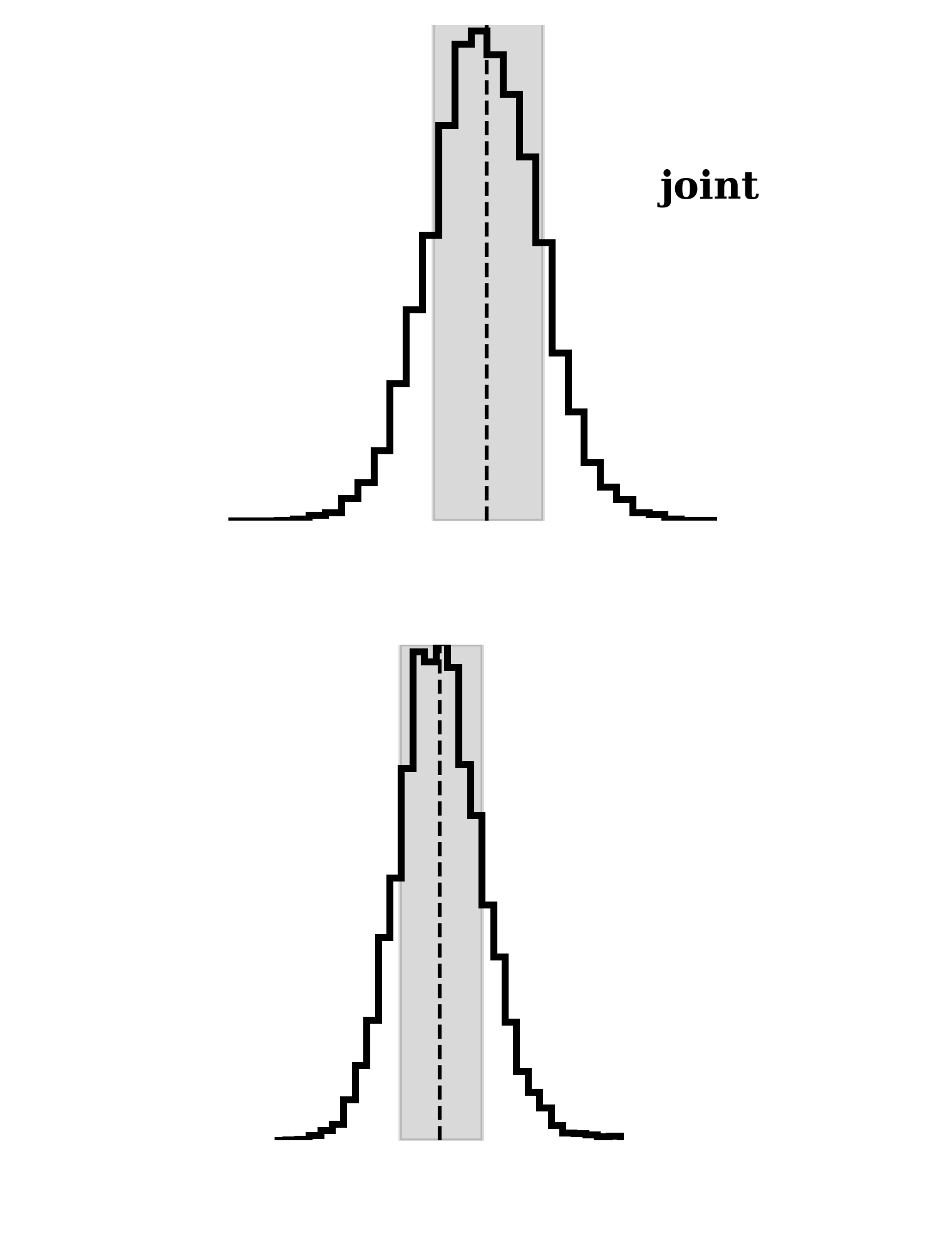}%
  \end{ocg}
  \hspace{-\hsize}%
  \caption{The 1D marginalized posterior PDFs of the K2-18b and c semi-amplitudes from analyses
    of the
    \ToggleLayer{fig:HKon,fig:HKoff}{\protect\cdbox{full HARPS (\emph{blue})}}, the
    \ToggleLayer{fig:CKon,fig:CKoff}{\protect\cdbox{reduced CARMENES (\emph{red})}}, and their
    \ToggleLayer{fig:jKon,fig:jKoff}{\protect\cdbox{joint (\emph{black})}} RV time-series. The
    \emph{dashed vertical lines} and \emph{shaded regions} depict the maximum a-posteriori values
    and $1\sigma$ confidence intervals respectively. 
    All $K_b$ and $K_c$ values are consistent at the $1\sigma$ level which is approximated by each
    PDF's 16$^{th}$ and 84$^{th}$ percentiles.}
  \label{fig:postK}
\end{figure}

In our re-analysis of the joint HARPS+CARMENES RVs we have measured the most likely keplerian solution to each planet's
orbit. By including all available RV observations of the K2-18 system (excluding those which are known to be anomalous),
we have obtained the most precise planetary solutions for K2-18 to-date.
The point estimates of the 2-planet model parameters resulting
from this analysis are presented in Table~\ref{table:k218}. As a sanity check we can compare the resulting marginalized
posterior PDFs for parameters of interest between the individual HARPS, CARMENES, and their joint RV time-series. In this way
we can ensure that the planetary solutions from the two spectrograph's time-series are consistent with each other 
as well as with their joint time-series. For instance, we compare the resulting marginalized poster PDFs of $K_b$ and
$K_c$ obtained with each time-series in Fig.~\ref{fig:postK}. It is evident that the MAP $K_b$ solutions are nearly
equivalent when measured with any of the three time-series. Similarly, MAP $K_c$ values are consistent at the $1\sigma$
level, albeit with more dispersion than the $K_b$ PDFs given the comparatively larger uncertainties in the K2-18c
ephemeris.

\subsection{Improved stellar parameters based on GAIA DR2}
To map the observable transit and RV parameters to physical planetary parameters we must first characterize the
host star. Specifically, we can exploit the exquisite precision of the GAIA DR2 to improve the stellar mass and
radius of K2-18.

Firstly, the K2-18 stellar mass is computed from the M dwarf mass-luminosity relation (MLR) from \cite{benedict16}.
The analytical MLR based on absolute $K$-band magnitudes is favoured over the $V$-band whose dispersion about the
relation is twice that in the $K$-band. The distance modulus is calculated from the precision GAIA DR2 stellar parallax
\citep[$p=26.299\pm 0.055$ mas;][]{gaia18} to be $\mu=2.900\pm 0.005$ where we have added the 30 $\mu$as systematic
offset in the measured parallax as noted in \cite{lindegren18}. By propagating errors
in the K2-18 $K$-band magnitude \citep[$K=8.899\pm 0.019$;][]{cutri03}, the distance modulus, and the MLR coefficients,
we find an absolute $K$-band
magnitude of $M_{\text{K}}=5.999\pm 0.020$ and a corresponding stellar mass of $M_s=0.495\pm 0.004$ M$_{\odot}$.

From the stellar mass we are able to derive the stellar radius using the empirical mass-radius relationship (MRR)
for M dwarfs from \cite{boyajian12}. By propagating the uncertainties in the M dwarf MRR
coefficients we compute the K2-18 stellar radius to be $R_s=0.469\pm 0.010$ R$_{\odot}$. We note that both the
updated stellar mass and radius---based on the stellar parallax---are considerably larger than the
spectroscopically-derived values of $0.359\pm 0.047$ M$_{\odot}$ and $0.411\pm 0.038$ $R_{\odot}$
\citep{benneke17}. The new mass and radius values are inconsistent with their previous values at the levels of
$2.9\sigma$ and $1.5\sigma$ respectively. This is the direct result of the increased K2-18 distance from GAIA 
(i.e. $38.025\pm 0.079$ pc) compared to its previously measured distance (i.e. $34\pm 4$ pc) and 
will have important implications for the derived physical parameters of
both K2-18b and c. We also note the improved fractional uncertainties on the updated stellar mass and radius
of 0.8\% and 2.1\% respectively compared to the previous fractional uncertainties of
13.1\% and 9.2\%.

\subsection{Precise planetary parameters}
The improved stellar parameters---along with our joint HARPS+CARMENES RV analysis---provide the most
precise set of planetary parameters for the planets K2-18b and c to-date. Point estimates of the planetary
parameters from our joint HARPS+CARMENES RV analysis are presented in Table~\ref{table:k218}. In
particular, we measure the precise mass and minimum mass of K2-18b and c respectively to be
$m_{p,b}=8.64\pm 1.35$ M$_{\oplus}$ and $m_{p,c}\sin{i_c}=5.63\pm 0.84$ M$_{\oplus}$.

The improved stellar radius also provides a more precise planetary radius given the measured $r_{p,b}/R_s$
value from \cite{benneke17}.
We find that $r_{p,b}=2.711 \pm 0.065$ R$_{\oplus}$. From this we derive a planetary
bulk density for K2-18b of $\rho_{p,b}=2.4\pm 0.4$ g cm$^{-3}$ thus making K2-18b inconsistent with either an
Earth-like composition or a pure water-world \citep{zeng13}. Prior to updating mass and radius of K2-18b,
neither of these scenarios could have been ruled out. It is now clear that at minimum, $\sim 8$\% of the size of
K2-18b (i.e. $\sim 1382$ km) must be attributed to an optically think gaseous atmosphere as evidenced by its
low bulk density. The expected signal amplitude in transmission for a cloud-free hydrogen-dominated atmosphere
($\mu=2$) is $\sim 10Hr_{p,b}/R_s^2\sim 155$ ppm
where $H=k_{\text{B}}T_{\text{eq}}/\mu m_{\text{p}} g$ is the atmospheric pressure scale
height, $k_{\text{B}}$ is the Boltzmann constant, $T_{\text{eq}}$ is the planet's equilibrium temperature assuming
an Earth-like Bond albedo, $\mu m_{\text{p}}$ is the assumed mean molecular weight, and $g$ is the surface gravity 
\citep{kaltenegger09}. For comparison, a well-mixed water-dominated atmosphere ($\mu=18$) has a transmission signal
amplitude of $\sim 18$ ppm. Given the scale height of its extended gaseous envelop and its proximity to the Solar
System, K2-18b continues to represent an exciting opportunity
to characterize a sub-Neptune-sized exoplanet receiving Earth-like insolation with upcoming space missions such
as the James Webb Space Telescope and ARIEL.

\begin{table*}
\small
\renewcommand{\arraystretch}{0.7}
\centering
\caption[]{K2-18 model parameters from the HARPS+CARMENES joint RV analysis}
\label{table:k218}
\begin{tabular}{lc}
\hline \\ [-1ex]
Parameter & Point estimate \smallskip\\
\hline \\ [-1ex]
\emph{Stellar Parameters} & \smallskip \\
2MASS Photometry & $J$=9.763$\pm$0.028, $H$=9.135$\pm$0.026, $K_s$=8.899$\pm$0.019 \\
Stellar mass, $M_s$ [M$_{\odot}$]  &  $0.495\pm 0.004$ \\
Stellar radius, $R_s$ [R$_{\odot}$]  &  $0.469\pm 0.010$  \\
Effective temperature, $T_{\text{eff}}$ [K]  & 3503 $\pm$ 60 \\
Stellar parallax, $p$ [mas] & $26.299\pm 0.055$ \\
Distance, $d$ [pc] & $38.025 \pm 0.079$ \\
HARPS systemic velocity, $\gamma_{0,\text{HARPS}}$ [m s$^{-1}$] & $652.51 \pm 1.0$ \\
CARMENES systemic velocity, $\gamma_{0,\text{CARMENES}}$ [m s$^{-1}$] & $-2.87\pm 0.9$ \medskip \\
\emph{GP hyperparameters} & \smallskip \\
HARPS covariance amplitude, $a_{\text{HARPS}}$ [\mps{]} & $3.0^{+3.4}_{-1.7}$  \\
CARMENES covariance amplitude, $a_{\text{CARMENES}}$ [\mps{]} & $5.0^{+5.3}_{-2.9}$ \\
Exponential timescale, $\lambda$ [days] &  $448.8\pm 67.3$ \\
Coherence, $\Gamma$ &  $0.17^{+0.07}_{-0.04}$ \\
Periodic timescale, $P_{\text{GP}}$ [days] & $37.4^{+0.5}_{-0.3}$ \\
HARPS additive jitter, $s_{\text{HARPS}}$ [\mps{]} & $0.48\pm 0.42$ \\
CARMENES additive jitter, $s_{\text{CARMENES}}$ [\mps{]} & $0.58\pm 0.53$ \medskip \\
\emph{K2-18c} & \smallskip \\
Period, $P_c$ [days] & $8.997 \pm 0.007$  \\
Time of inferior conjunction, $T_{0,c}$ [BJD-2,450,000] & 7263.69 $\pm$ 0.44 \\
Radial velocity semi-amplitude, $K_c$ [\mps{]} & $2.76 \pm 0.41$ \\
$h_c =\sqrt{e_c}\cos{\omega_c}$ & $0.00^{+0.24}_{-0.30}$ \\
$k_c =\sqrt{e_c}\sin{\omega_c}$ & $0.15^{+0.23}_{-0.28}$ \\
Semi-major axis, $a_c$ [AU] & $0.0670 \pm 0.0002$ \\
Minimum planet mass, $m_{p,c} \sin{i_c}$ [M$_{\oplus}$] & $5.62 \pm 0.84$ \\
Equilibrium temperature, $T_{\text{eq},c}$ [K] & \\
\hspace{2pt} Bond albedo of 0.3 & 409 $\pm$ 8 \medskip \\
\emph{K2-18b} & \smallskip \\
Period, $P_b$ [days] & $32.93962 \pm 1.0 \times 10^{-4}$ \\
Time of inferior conjunction, $T_{0,b}$ [BJD-2,450,000] & $7264.39142 \pm 6.4 \times 10^{-4}$ \\
Radial velocity semi-amplitude, $K_b$ [\mps{]} & $2.75 \pm 0.43$ \\
$h_b =\sqrt{e_b}\cos{\omega_b}$ & $0.30^{+0.11}_{-0.24}$ \\
$k_b =\sqrt{e_b}\sin{\omega_b}$ & $-0.05^{+0.26}_{-0.25}$ \\
Semi-major axis, $a_b$ [AU] & $0.1591 \pm 0.0004$ \\
Planet radius, $r_{p,b}$ [R$_{\oplus}$]$^{(\bullet)}$ & $2.711\pm 0.065$ \\
Planet mass, $m_{p,b}$ [M$_{\oplus}$]$^{(\ast)}$ & $8.63 \pm 1.35$ \\
Planet density, $\rho_{p,b}$ [$\mathrm{g\;cm^{-3}}$] & $2.4\pm 0.4$ \\
Surface gravity, $g$ [$\mathrm{m\;s^{-2}}$]  & $11.5\pm 1.9$ \\
Escape velocity, $v_{\text{esc},b}$ [$\mathrm{km\;s^{-1}}$] & $19.9\pm 1.6$ \\
Equilibrium temperature, $T_{\text{eq},b}$ [K] & \\
\hspace{2pt} Bond albedo of 0.3 & 265 $\pm$ 5 \medskip \\
\hline
\end{tabular}
\begin{list}{}{}
\item {\bf{Notes.}}
  $^{(\bullet)}$ based on the measured $r_{p,b}/R_s$ of K2-18b from \cite{benneke17}; $r_{p,b}/R_s=0.05295\pm 0.00060$. \\
  $^{(\ast)}$ assuming the measured orbital inclination of K2-18b from \cite{benneke17}; $i_b = 89.5785^{+0.0079}_{-0.0088}$ degrees. \\
\end{list}
\end{table*}

\appendix
\section{Iterative radial velocity time-series and GLSP figures from Sect.~\ref{sect:correlated}}
In Sect.~\ref{sect:correlated} we considered a variety of RV datasets and models which included either
1 or 2 planets along with a GP regression model of stellar activity that had been trained on the star's
K2 photometry. The following figures depict the iterative RVs and GLSPs for each dataset and model. In
each iteration we remove one or more coherent signals (i.e. planets or activity) to see if any residual
periodicities persist for which additional model components may be required.

\begin{figure*} 
  \centering
  \includegraphics[width=0.77\hsize]{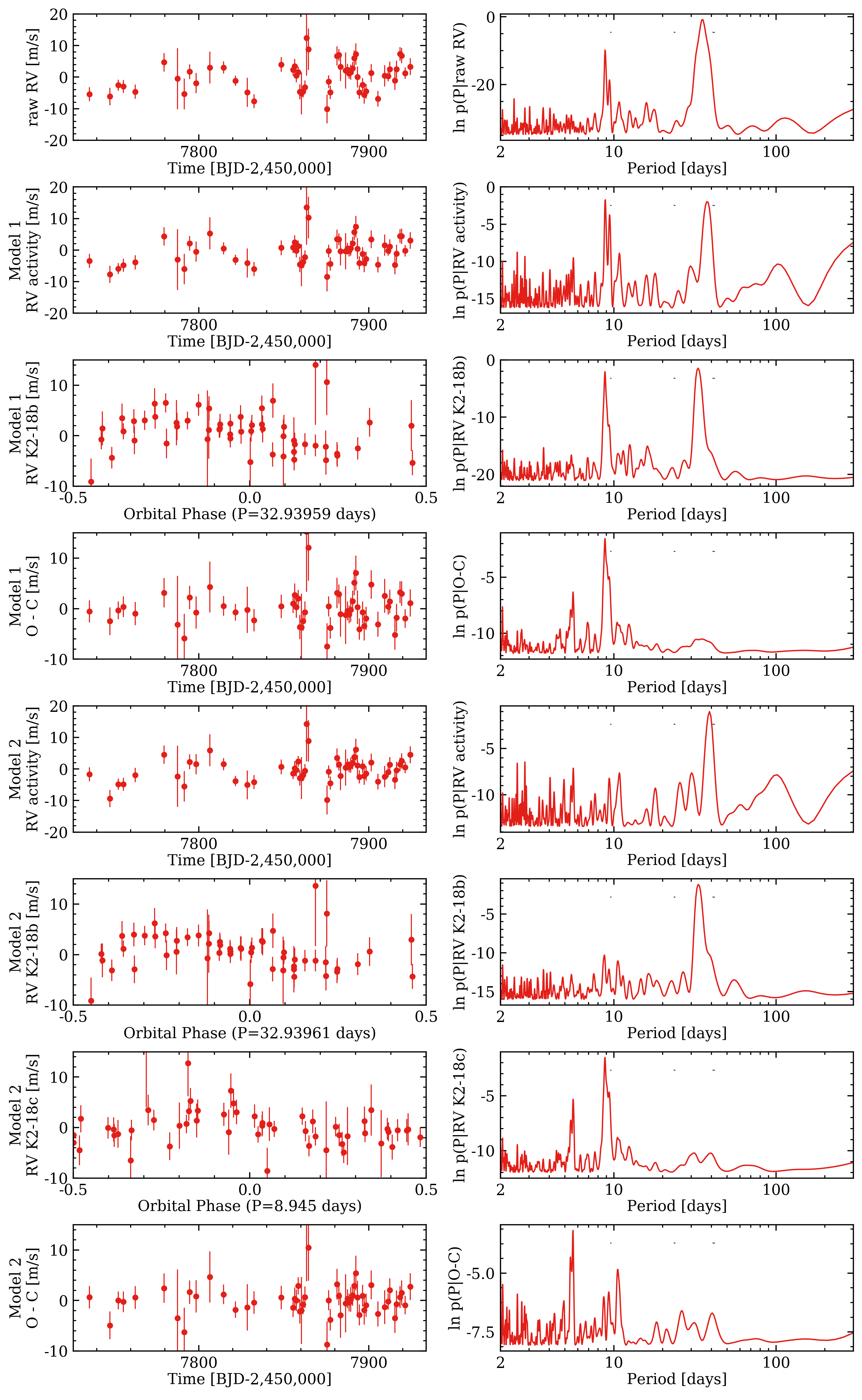}%
  \hspace{-0.77\hsize}%
  \begin{ocg}{fig:MCoff}{fig:MCoff}{0}%
  \end{ocg}%
  \begin{ocg}{fig:MCon}{fig:MCon}{1}%
  \includegraphics[width=0.77\hsize]{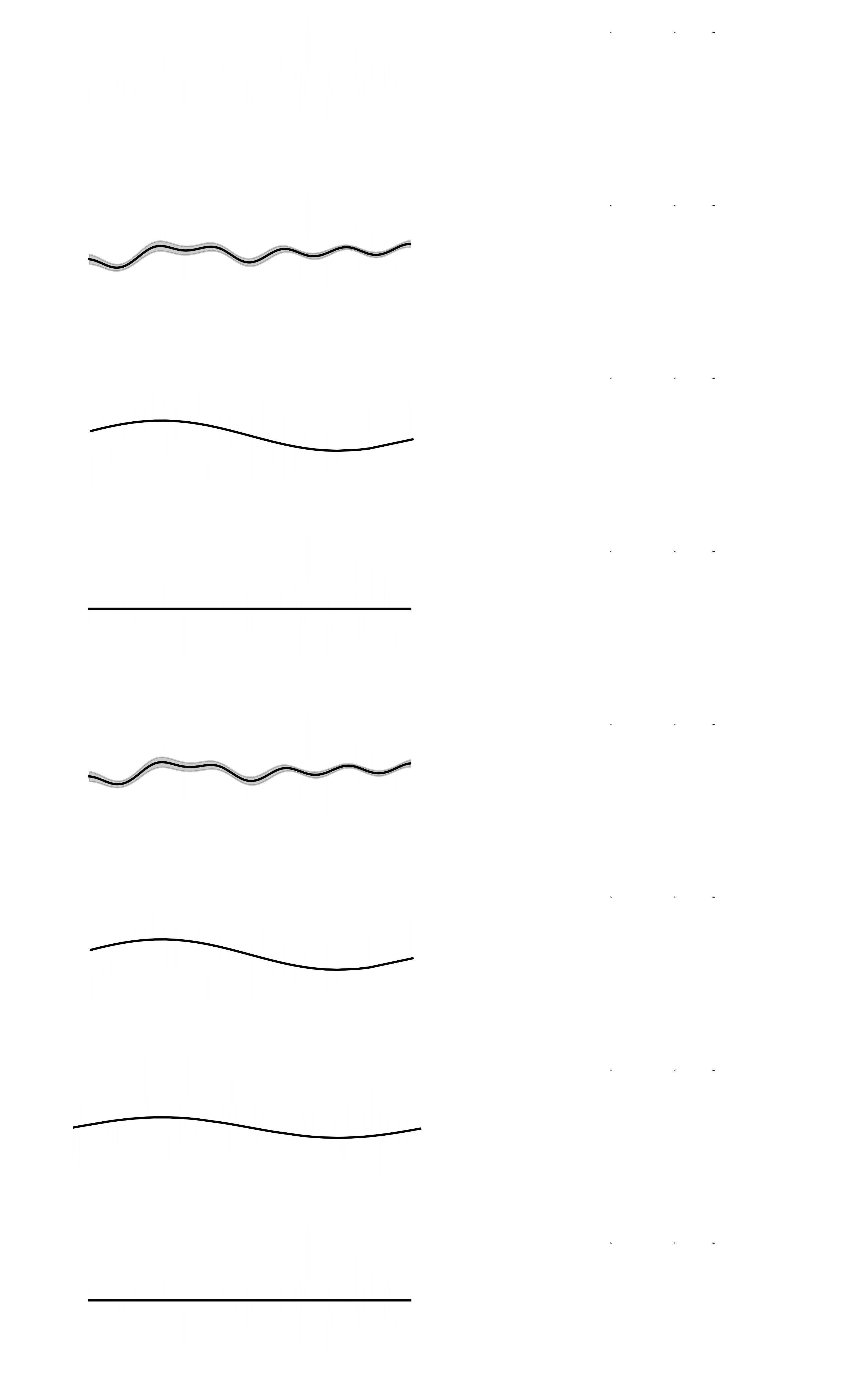}%
  \end{ocg}
  \hspace{-0.77\hsize}%
  \begin{ocg}{fig:LCoff}{fig:LCoff}{0}%
  \end{ocg}%
  \begin{ocg}{fig:LCon}{fig:LCon}{1}%
  \includegraphics[width=0.77\hsize]{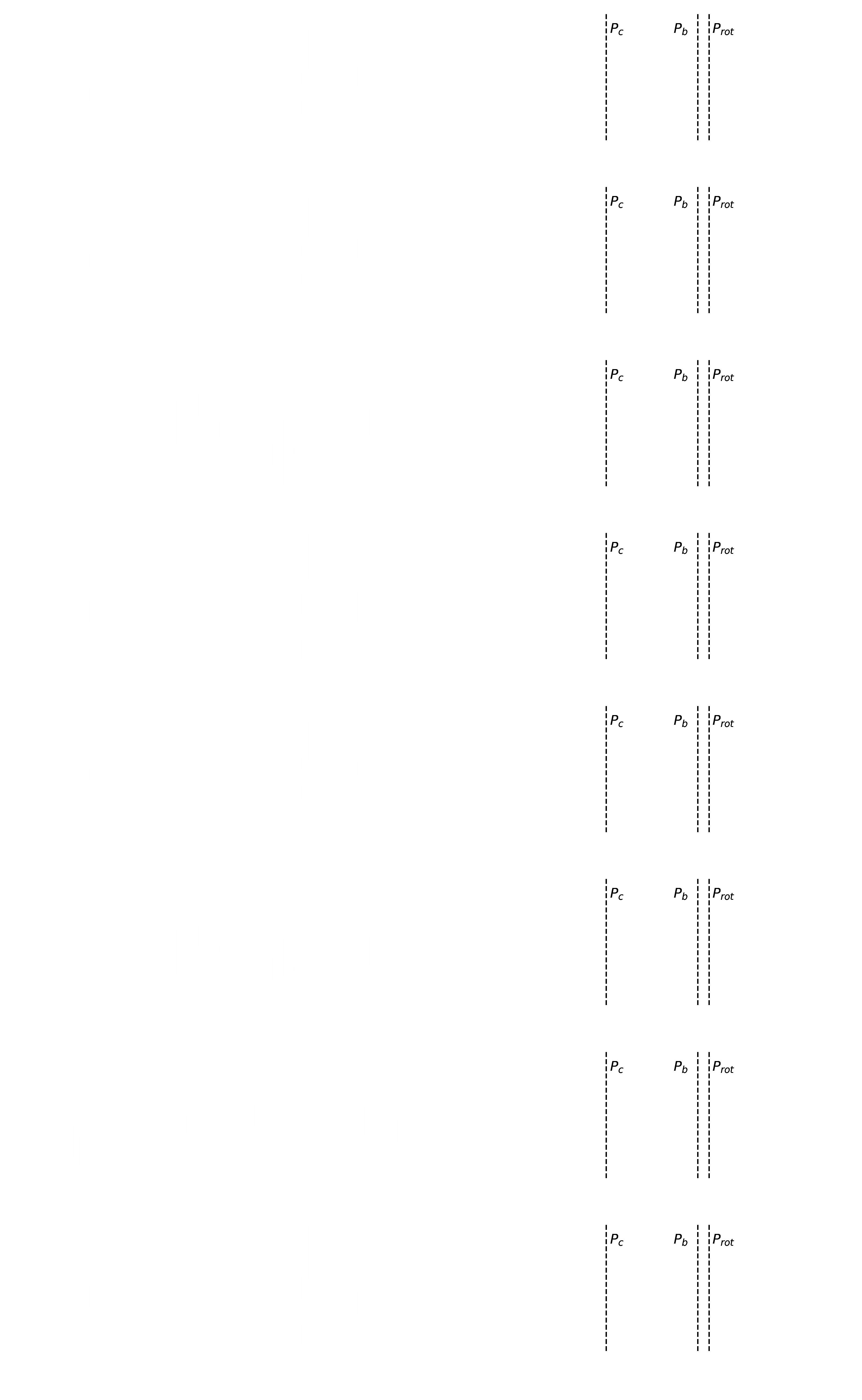}%
  \end{ocg}
  \hspace{-0.77\hsize}%
  \caption{Results of our RV analysis of the 55 CARMENES RVs that are known not to significantly
    suppress the apparent 9 day signal seen with HARPS. The RV time-series and their corresponding GLSP
    are plotted in common rows for each coherent RV signal modelled (i.e. planets and stellar
    activity) in either a 1 or 2-planet model. The over-plotted
    \ToggleLayer{fig:MCon,fig:MCoff}{\protect\cdbox{RV models}} are computed using
    the MAP model parameters from our MCMC analysis. The
    \ToggleLayer{fig:LCon,fig:LCoff}{\protect\cdbox{\emph{vertical dashed lines}}} in the GLSPs are
    indicative of the MAP orbital periods for K2-18b and c and the photometric stellar rotation period. 
    The first row depicts the raw RVs, the three proceeding rows present the results
    assuming a 1-planet model (i.e. K2-18b), and the final four rows present the results assuming a 2-planet
    model (i.e. K2-81b and c). The residual rms values assuming a 1 and 2-planet model are
    3.84 and 3.57 \mps{} respectively. We find that the source of the residual $\sim 5.5$ day signal in the
    bottom GLSP is an alias rather than being due to an unmodelled physical source (see text).}
  \label{fig:analysisC}
\end{figure*}

\begin{figure*} 
  \centering
  \includegraphics[width=0.77\hsize]{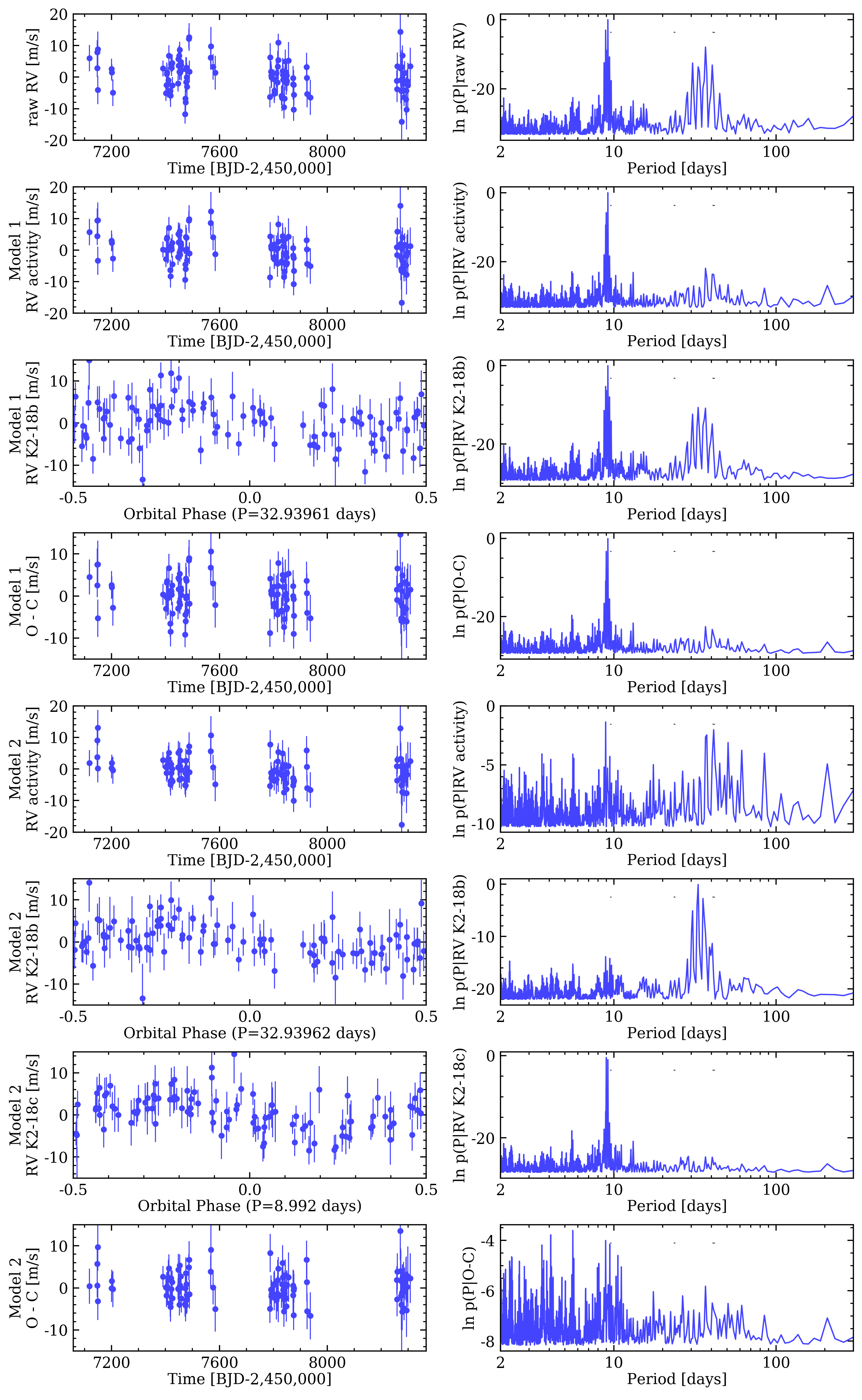}%
  \hspace{-0.77\hsize}%
  \begin{ocg}{fig:MH106off}{fig:MH106off}{0}%
  \end{ocg}%
  \begin{ocg}{fig:MH106on}{fig:MH106on}{1}%
  \includegraphics[width=0.77\hsize]{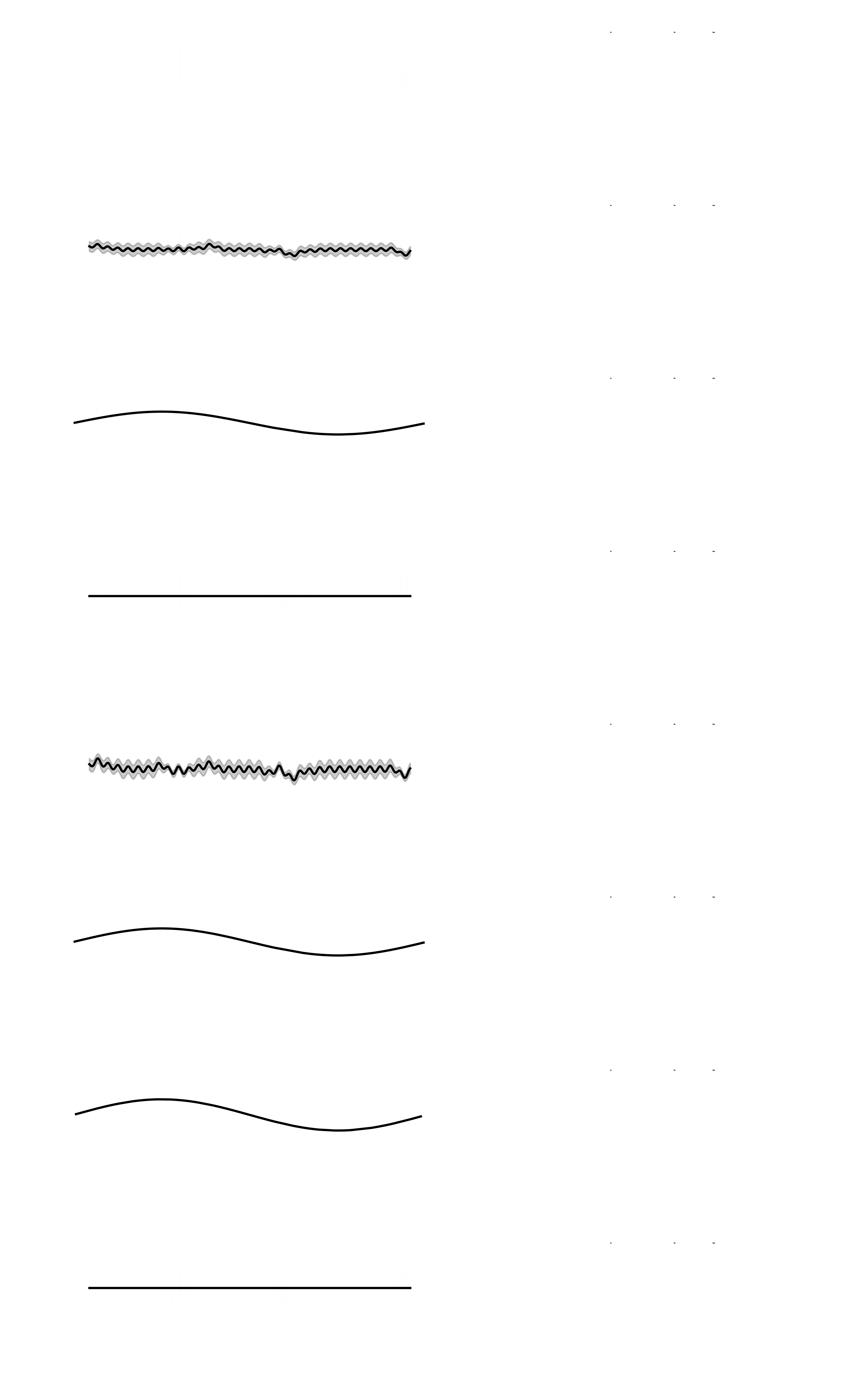}%
  \end{ocg}
  \hspace{-0.77\hsize}%
  \begin{ocg}{fig:LH106off}{fig:LH106off}{0}%
  \end{ocg}%
  \begin{ocg}{fig:LH106on}{fig:LH106on}{1}%
  \includegraphics[width=0.77\hsize]{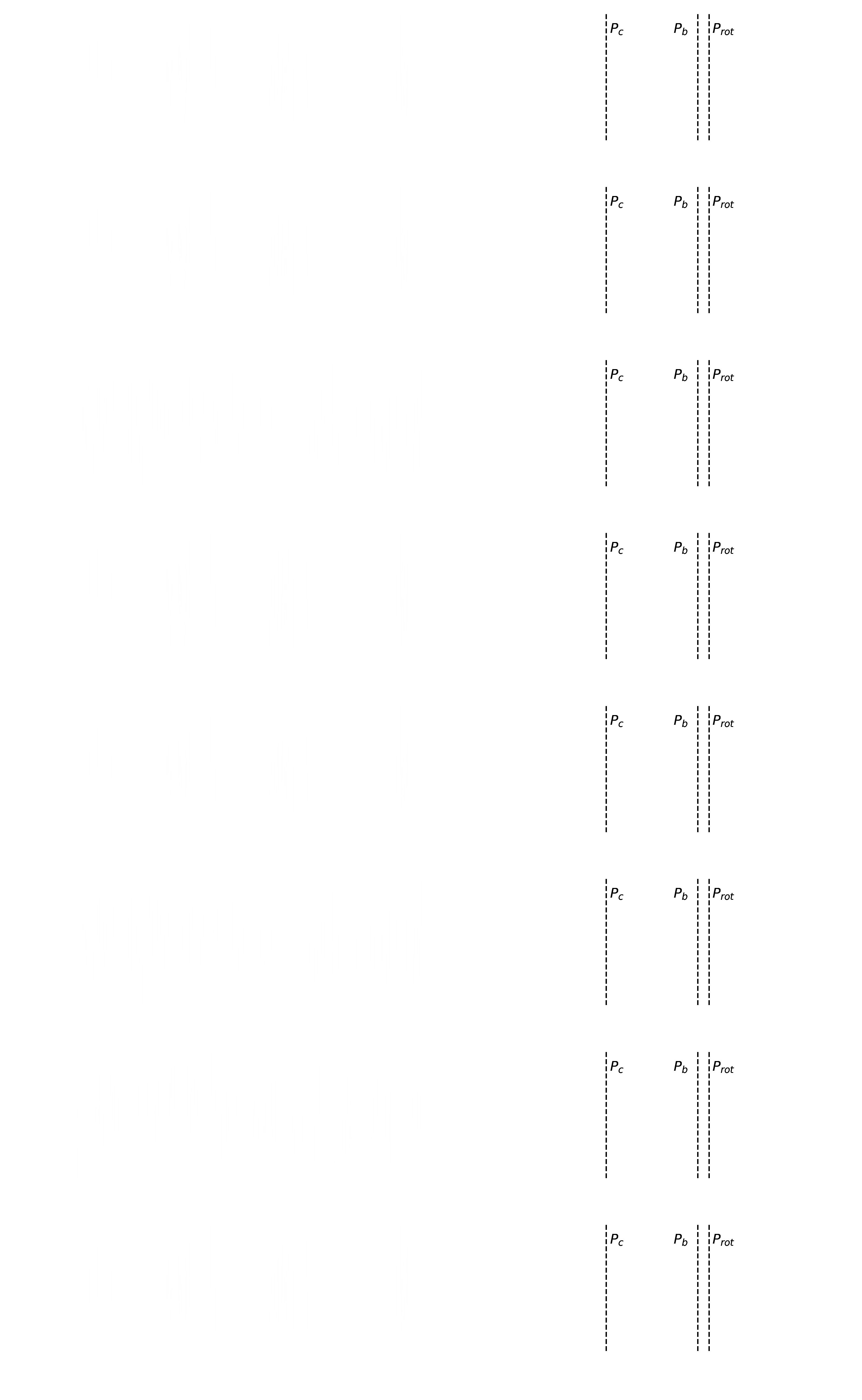}%
  \end{ocg}
  \hspace{-0.77\hsize}%
  \caption{Similar to Fig.~\ref{fig:analysisC} but for the full HARPS WF containing 106 RVs.
    The rms of the residual time-series assuming a 1 and 2-planet model are
    4.68 and 3.93 \mps{} respectively.
    \ToggleLayer{fig:MH106on,fig:MH106off}{\protect\cdbox{MAP RV models}}.
    \ToggleLayer{fig:LH106on,fig:LH106off}{\protect\cdbox{GLSP periodicities}}.}
  \label{fig:analysisH106}
\end{figure*}

\begin{figure*} 
  \centering
  \includegraphics[width=0.77\hsize]{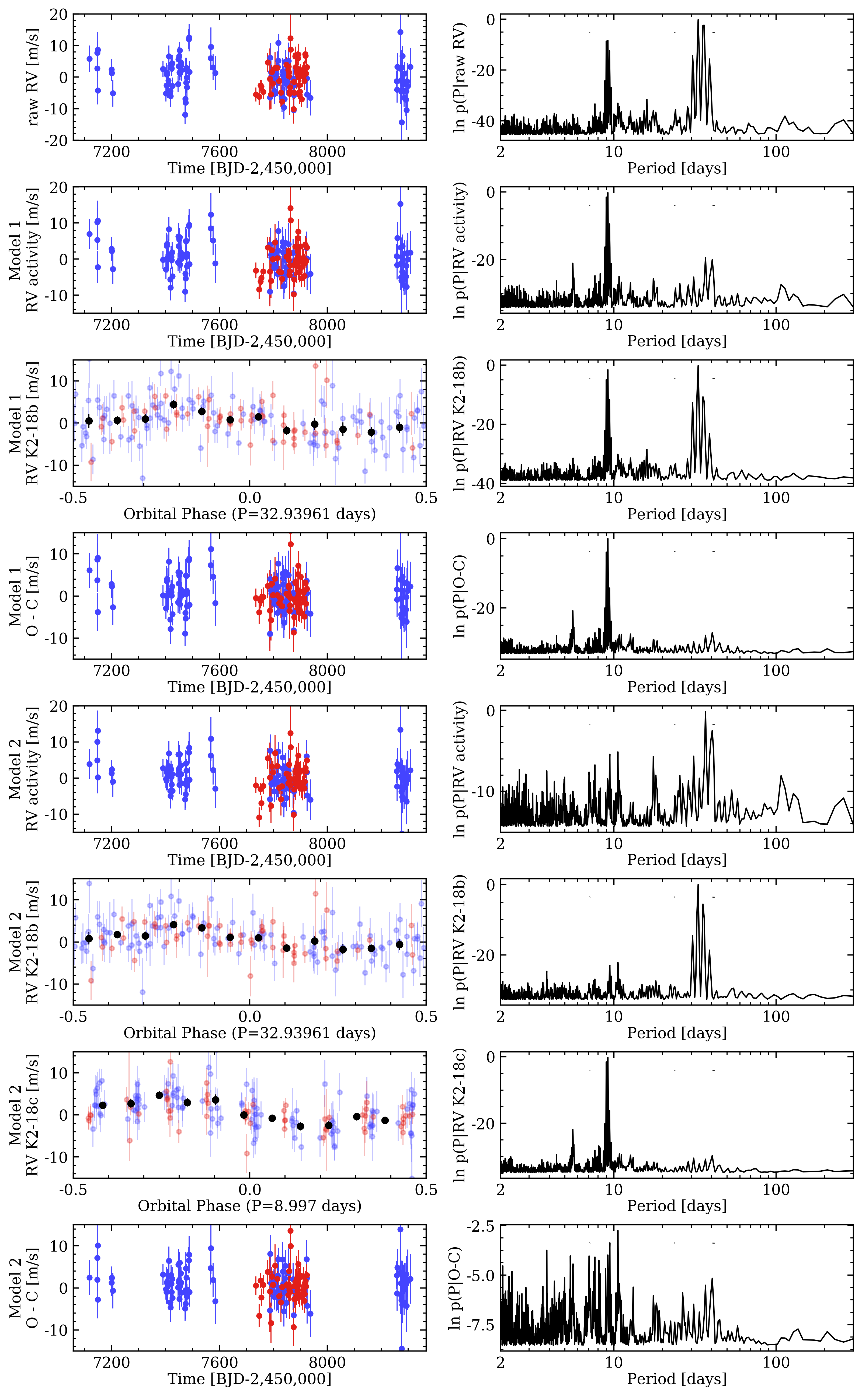}%
  \hspace{-0.77\hsize}%
  \begin{ocg}{fig:MHCoff}{fig:MHCoff}{0}%
  \end{ocg}%
  \begin{ocg}{fig:MHCon}{fig:MHCon}{1}%
  \includegraphics[width=0.77\hsize]{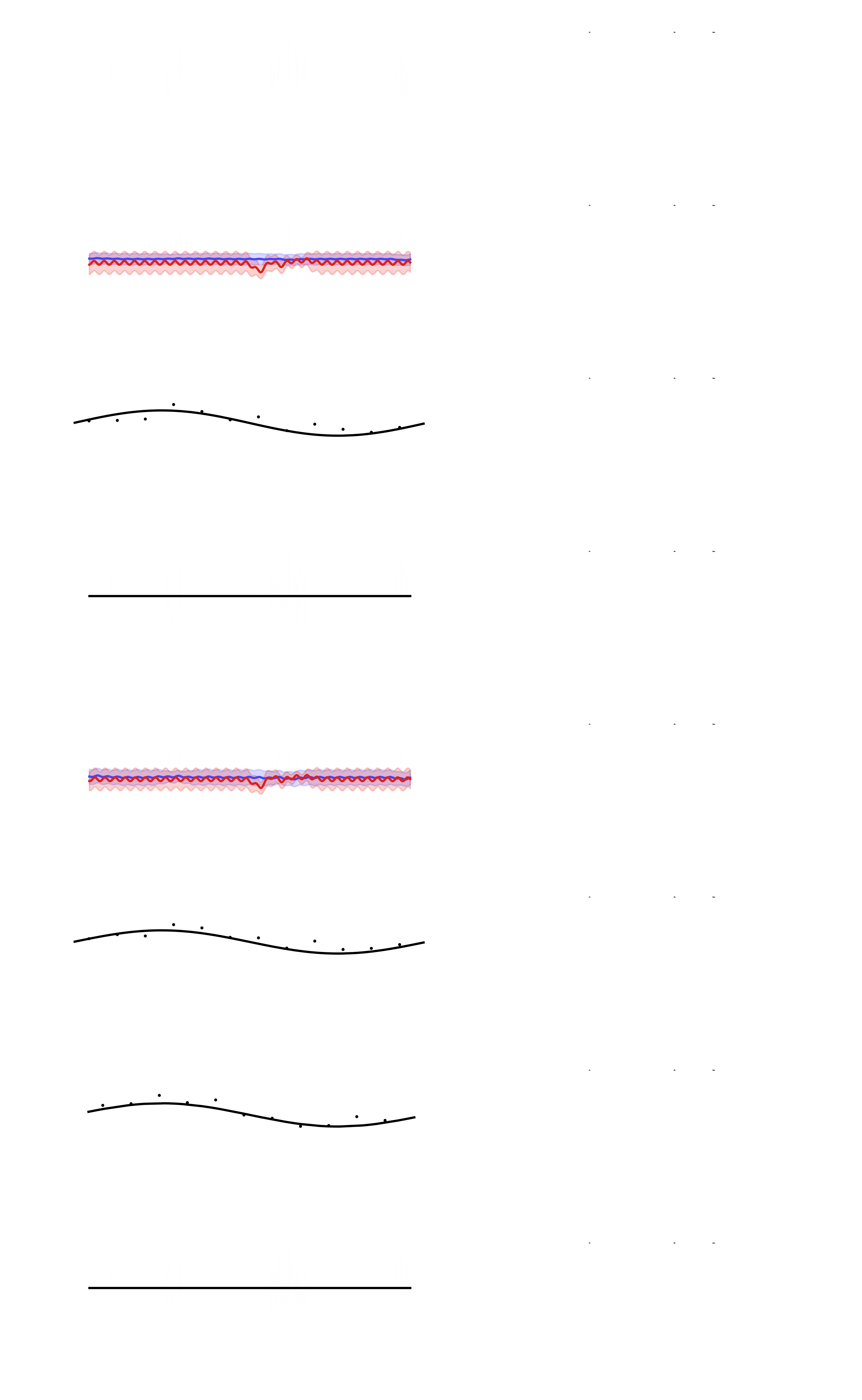}%
  \end{ocg}
  \hspace{-0.77\hsize}%
  \begin{ocg}{fig:LHCoff}{fig:LHCoff}{0}%
  \end{ocg}%
  \begin{ocg}{fig:LHCon}{fig:LHCon}{1}%
  \includegraphics[width=0.77\hsize]{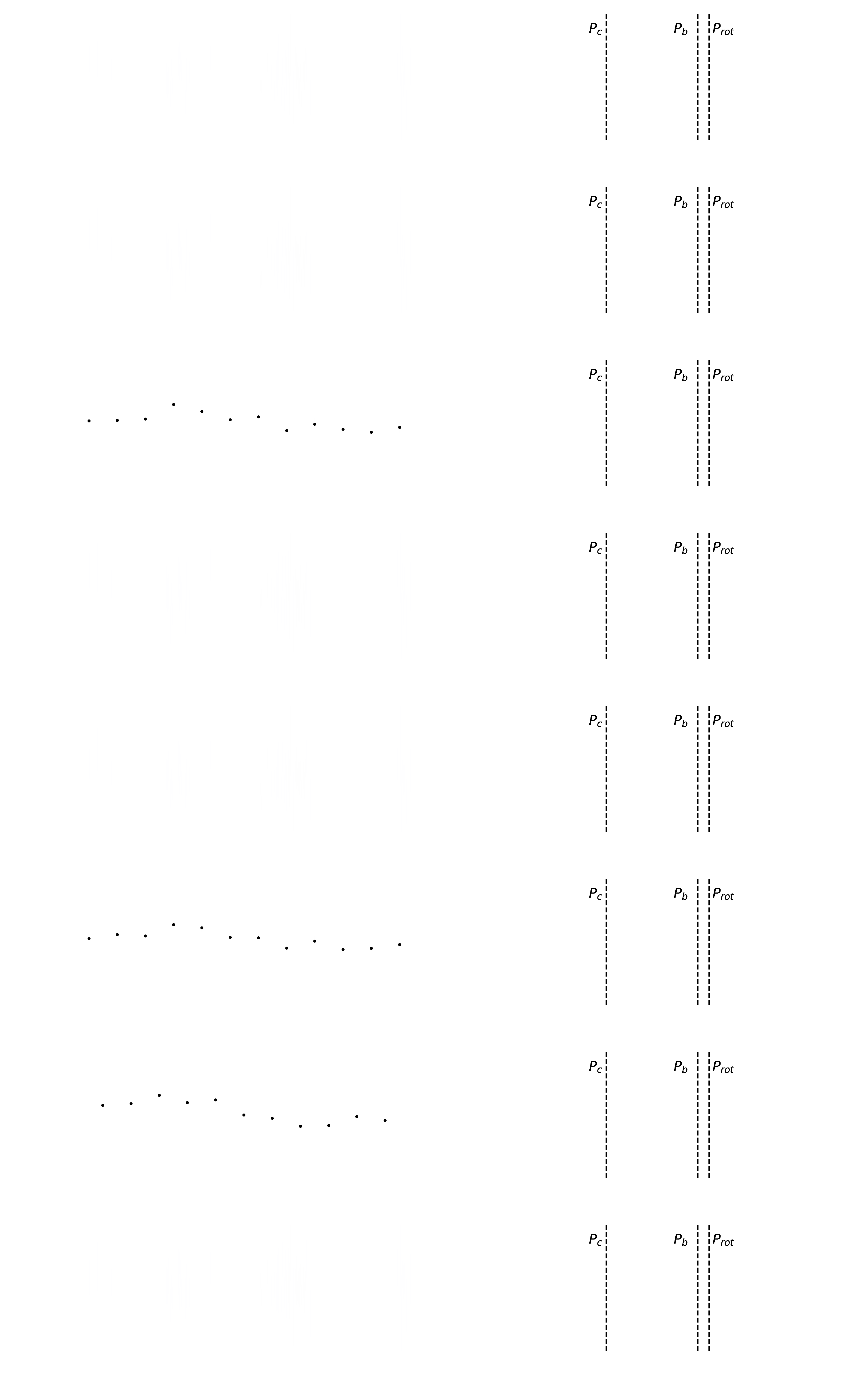}%
  \end{ocg}
  \hspace{-0.77\hsize}%
  \caption{Similar to Fig.~\ref{fig:analysisC} but for the 161 joint HARPS+CARMENES RVs.
    The HARPS and CARMENES RVs are plotted as \emph{blue} and \emph{red} markers respectively.
    The phase-folded RVs depicting planetary signals are binned for clarity.
    The rms of the residual time-series assuming a 1 and 2-planet model are
    4.51 and 3.82 \mps{} respectively.
    \ToggleLayer{fig:MHCon,fig:MHCoff}{\protect\cdbox{MAP RV models}}.
    \ToggleLayer{fig:LHCon,fig:LHCoff}{\protect\cdbox{GLSP periodicities}}.}
  \label{fig:analysisHC}
\end{figure*}

\begin{figure*} 
  \centering
  \includegraphics[width=0.77\hsize]{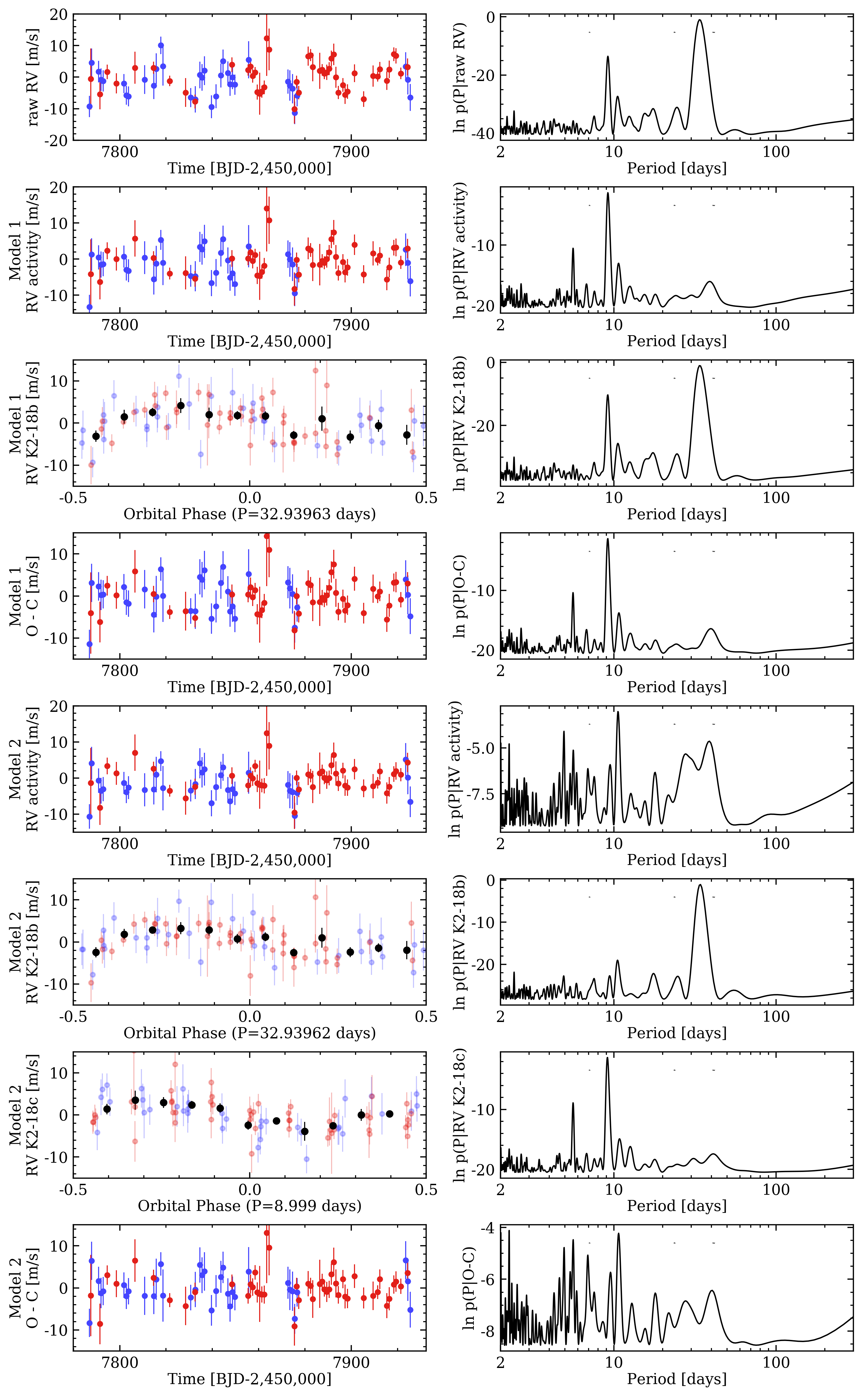}%
  \hspace{-0.77\hsize}%
  \begin{ocg}{fig:MHCOoff}{fig:MHCOoff}{0}%
  \end{ocg}%
  \begin{ocg}{fig:MHCOon}{fig:MHCOon}{1}%
  \includegraphics[width=0.77\hsize]{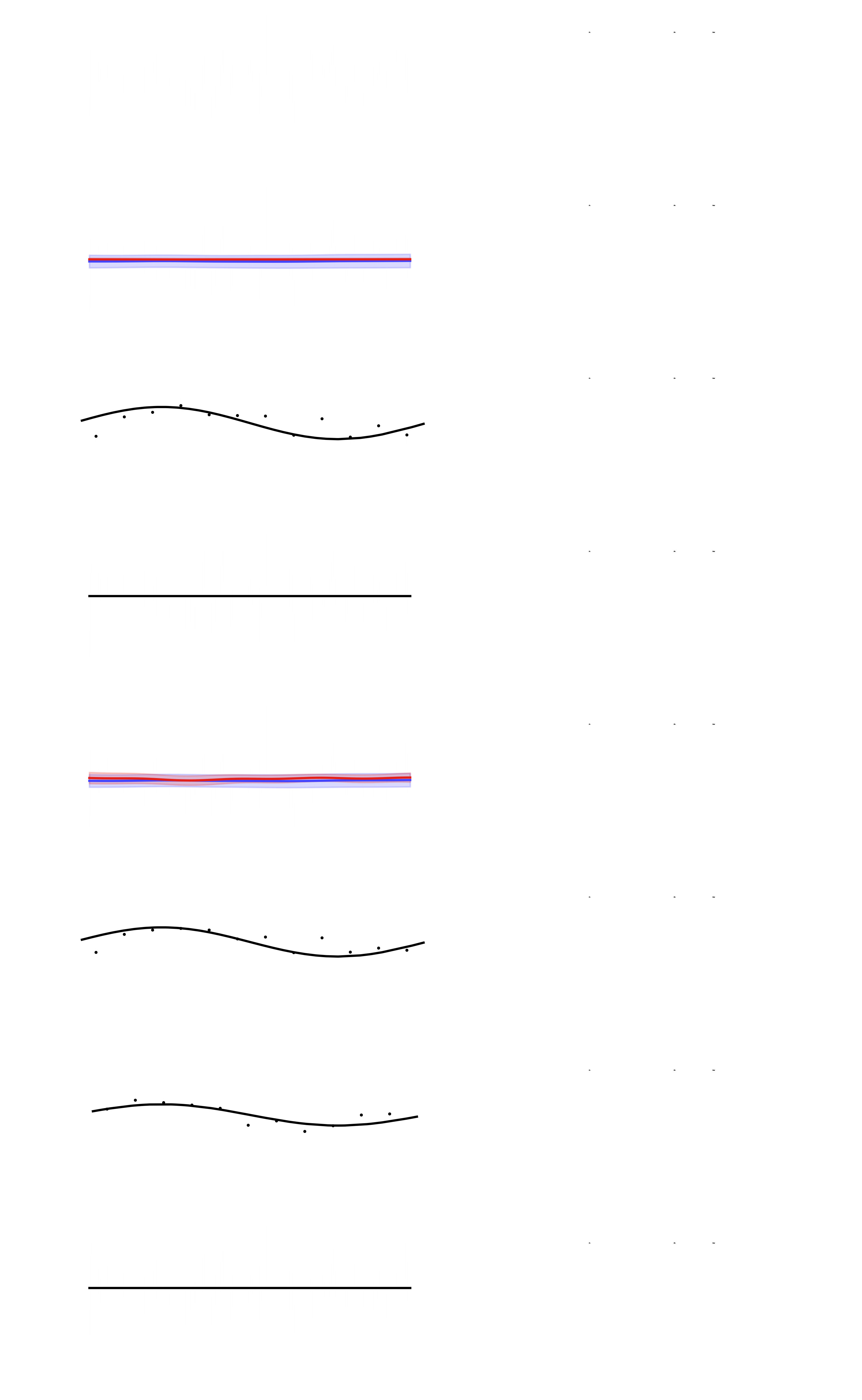}%
  \end{ocg}
  \hspace{-0.77\hsize}%
  \begin{ocg}{fig:LHCOoff}{fig:LHCOoff}{0}%
  \end{ocg}%
  \begin{ocg}{fig:LHCOon}{fig:LHCOon}{1}%
  \includegraphics[width=0.77\hsize]{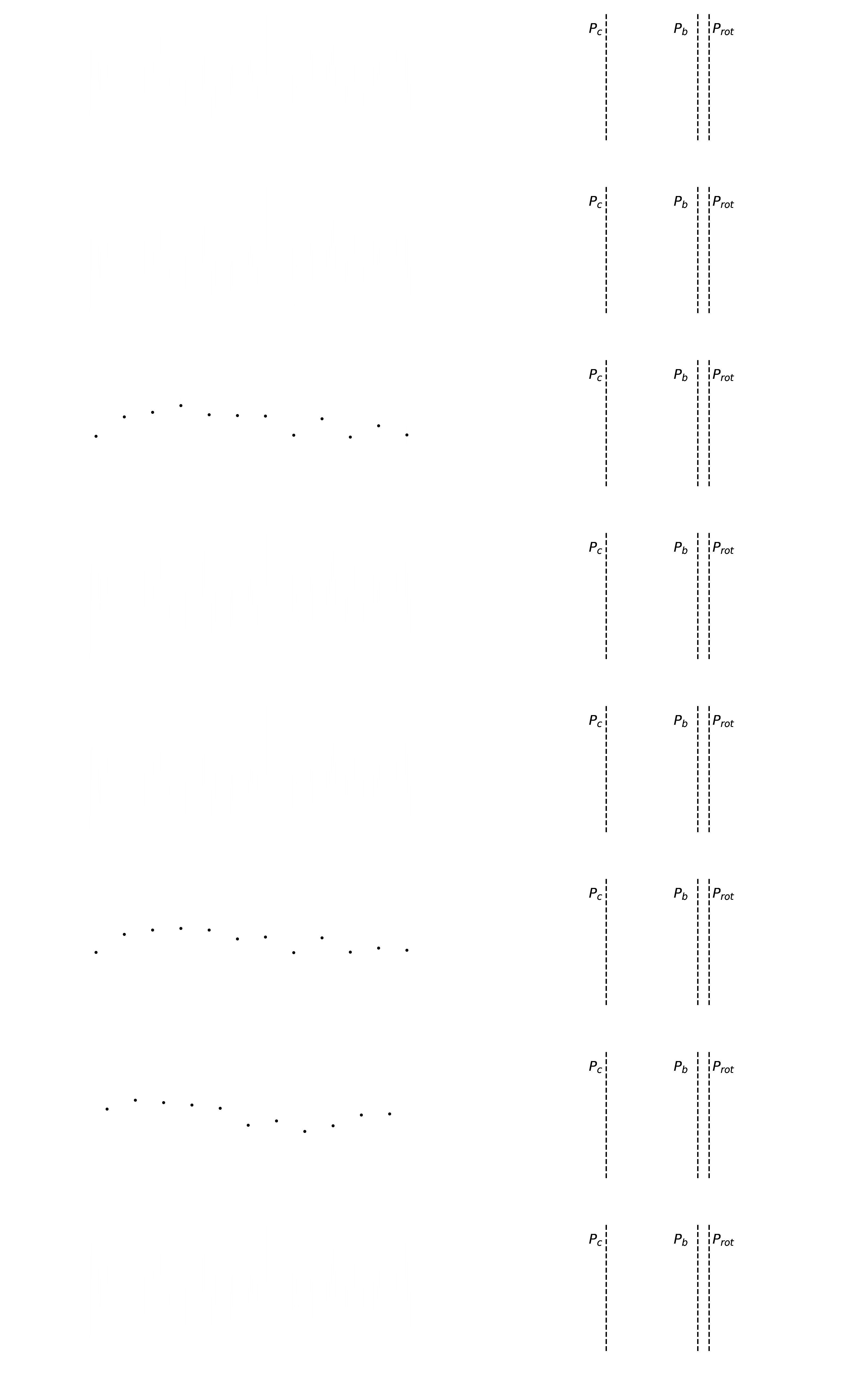}%
  \end{ocg}
  \hspace{-0.77\hsize}%
  \caption{Similar to Fig.~\ref{fig:analysisC} but for the 84 joint HARPS+CARMENES RVs
    which were obtained during the time interval in which the two spectrograph WFs overlap; i.e.
    from February-June 2017.
    The HARPS and CARMENES RVs are plotted as \emph{blue} and \emph{red} markers respectively.
    The phase-folded RVs depicting planetary signals are binned for clarity.
    The rms of the residual time-series assuming a 1 and 2-planet model are
    4.26 and 3.62 \mps{} respectively.
    \ToggleLayer{fig:MHCOon,fig:MHCOoff}{\protect\cdbox{MAP RV models}}.
    \ToggleLayer{fig:LHCOon,fig:LHCOoff}{\protect\cdbox{GLSP periodicities}}.}
  \label{fig:analysisHCoverlap}
\end{figure*}

\begin{acknowledgements}
  R.C. thanks the anonymous referee and the CARMENES team---particularly P. Sarkis and T. Henning---for taking
  the time to read and comment on the manuscript during its editing phase and particularly to the CARMENES team
  for investigating potential causes of the anomalous CARMENES RVs.
  R.C. and K.M. acknowledge support for this work from the Natural Sciences and Engineering Research Council of Canada.
  N.A.D. acknowledges support from FONDECYT 3180063.
  X.B., J.M.A, and A.W. acknowledge funding from the European Research Council under the ERC Grant Agreement 
  n. 337591-ExTrA.
  N.C.S. acknowledges the funding by FEDER - Fundo Europeu de Desenvolvimento Regional funds through
  the COMPETE 2020 - Programa Operacional Competitividade e Internacionaliza\c{c}\~{a}o (POCI), and
  by Portuguese funds through FCT - Funda\c{c}\~{a}o para a Ci\^{e}ncia e a Tecnologia in the framework
  of the projects POCI-01-0145-FEDER-028953 and POCI-01-0145-FEDER-032113, as well as from FCT and
  FEDER through COMPETE2020 to grants UID/FIS/04434/2013 \& POCI-01-0145-FEDER-007672, PTDC/FIS-AST/1526/2014 \&
  POCI-01-0145-FEDER-016886 and PTDC/FIS-AST/7073/2014 \& POCI-01-0145-FEDER-016880. 
\end{acknowledgements}

\bibliographystyle{aa}
\bibliography{refs}

\begin{sidewaystable*}
\tiny
\renewcommand{\arraystretch}{0.7}
\centering
\caption[]{Full HARPS Time-Series from \citetalias{cloutier17b} and this work}
\label{table:data}
\begin{tabular}{ccccccccccccc}
\hline \\ [-1ex]
BJD-2,450,000 & RV & $\sigma$RV & blue RV & blue $\sigma$RV & red RV & red $\sigma$RV & NaD & $\sigma$NaD & H$\alpha$ & $\sigma$H$\alpha$ & FWHM & BIS \\
& [m s$^{-1}$] & [m s$^{-1}$] & [m s$^{-1}$] & [m s$^{-1}$] & [m s$^{-1}$] & [m s$^{-1}$] & & & &  & & \\
\hline \\ [-1ex]
7117.565870 & 659.11 & 4.16 & 670.79 & 7.12 & 652.95 & 5.75 & 0.01656 & 0.00044 & 0.06339 & 0.00039 & 3.067 & 4.750 \\
7146.526948 & 656.01 & 2.82 & 658.01 & 4.72 & 653.12 & 3.95 & 0.01275 & 0.00025 & 0.06649 & 0.00027 & 3.068 & -3.030 \\
7146.646070 & 660.97 & 3.85 & 654.96 & 6.68 & 661.33 & 5.28 & 0.01364 & 0.00038 & 0.06854 & 0.00037 & 3.079 & -6.310 \\
7148.518851 & 649.04 & 4.47 & 645.45 & 7.99 & 647.27 & 6.00 & 0.01429 & 0.00047 & 0.06665 & 0.00042 & 3.071 & 8.580 \\
7148.639664 & 661.91 & 5.62 & 684.89 & 10.16 & 652.86 & 7.44 & 0.01582 & 0.00064 & 0.06578 & 0.00052 & 3.069 & -12.900 \\
7199.503915 & 655.69 & 3.27 & 657.67 & 5.47 & 653.19 & 4.66 & 0.01290 & 0.00031 & 0.06653 & 0.00032 & 3.090 & 16.920 \\
7200.503114 & 654.59 & 2.65 & 657.05 & 4.44 & 657.69 & 3.77 & 0.01200 & 0.00023 & 0.06628 & 0.00025 & 3.080 & 19.340 \\
7204.491167 & 648.17 & 4.25 & 652.91 & 7.30 & 641.60 & 5.83 & 0.01257 & 0.00044 & 0.06411 & 0.00040 & 3.076 & -10.070 \\
7390.845075 & 655.87 & 2.53 & 651.28 & 3.92 & 658.23 & 3.85 & 0.01094 & 0.00022 & 0.06724 & 0.00026 & 3.106 & -0.200 \\
7401.779223 & 648.08 & 2.52 & 641.77 & 4.03 & 652.49 & 3.82 & 0.01041 & 0.00021 & 0.06675 & 0.00026 & 3.105 & 1.060 \\
\hline
\end{tabular}
\begin{list}{}{}
\item {\bf{Note.}}
  Only the first ten rows of this table are shown to demonstrate its format. The full time-series will be available in
  the online published version of this manuscript or by request to
  \href{mailto:cloutier@astro.utoronto.ca}{cloutier@astro.utoronto.ca}.
\end{list}
\end{sidewaystable*}

\end{document}